\documentclass[12pt]{iopart}
\usepackage{cite}

\expandafter\let\csname equation*\endcsname\relax
\expandafter\let\csname endequation*\endcsname\relax
\usepackage{amsmath}

\usepackage{graphicx, xcolor}
\usepackage{subcaption}

\begin{document}

\title{Optimization of stellarator configurations combining omnigenity and piecewise omnigenity}

\author{Hengqian Liu$^1$, Guodong Yu$^1$, Jos\'e Luis Velasco$^2$, Caoxiang Zhu$^1*$}

\address{$^1$  CAS Key Laboratory of Frontier Physics in Controlled Nuclear Fusion, School of Nuclear Science and Technology, University of Science and Technology of China, Hefei, Anhui 230026, China}
\address{$^2$Laboratorio Nacional de Fusión, CIEMAT, 28040 Madrid, Spain}
\ead{caoxiangzhu@ustc.edu.cn}
\vspace{10pt}

\begin{abstract}
We present a method for optimizing stellarator configurations that combine omnigenity and piecewise omnigenity (pwO).
Within the \texttt{OOPS} optimization framework [Liu \textit{et al.}, arXiv:2502.09350 (2025)], we introduce a mapping technique that can ``squeeze'' general omnigenous fields to approximate pwO in the high-field side.
Using this approach, we obtain a range of optimized configurations that combine poloidal omnigenity (PO) and pwO, spanning different field periods and aspect ratios.
We further show that these configurations are compatible with a magnetic well.
The resulting configurations exhibit favorable neoclassical transport and bootstrap current properties while partially relaxing the strict constraints of omnigenity.
These results suggest that such configurations are promising candidates for future stellarator reactors.
% A mapping method called `squeez' is introduced under the \texttt{OOPS} (Omnigenity OPtimization like quasiSymmetry) optimization framework \cite{liu_optimizing_2025}. By developing an omnigenity mapping near the high-field side, `squeez' approximates piecewise omnigenity (pwO) at this side while satisfying poloidal omnigenity (PO) at the low-field side, resulting in novel nearly-PO pwO configurations. Using this robust method, multiple configurations with diverse field periods and aspect ratios have been optimized. Their performance is evaluated regarding configuration complexity, neoclassical transport, and bootstrap current.
\end{abstract}

% \vspace{2pc}
% \noindent{\it Keywords}: Stellarator, Fusion，omnigenity, piecewise omnigenity
%\maketitle

\section{Introduction}
%General statements of stellarators, stellarator optimization
Stellarators confine plasmas with externally generated three-dimensional magnetic fields.
This eliminates the need for a current drive and avoids current-driven instabilities, enabling inherent steady-state plasma operation. However, the absence of axisymmetry in stellarators generally worsens particle confinement, and generic three-dimensional configurations exhibit large neoclassical transport unless specifically optimized.

% omnigenity
% Poor confinement is associated with trapped particles. In the collisionless limit, unlike passing particles (that remain confined to nested toroidal flux surfaces), trapped particles move back and forth along field lines while experiencing, on a longer time scale, guiding-center drifts with a radial component. To suppress the deleterious effect of these drifts, the magnetic field of the stellarator must be designed so that the time-averaged radial drift vanishes for every trapped orbit—a property known as omnigenity\cite{cary_helical_1997}. Omnigenous fields satisfy several stringent geometric criteria, including: contours of constant magnetic-field strength $|\mathbf{B}|$ that close poloidally, toroidally, or helically (in particular, straight contours of $B_{\max}$ in Boozer coordinates); and a bounce distance along a field line that is uniform in the field-line label. Optimization studies have produced configurations that closely satisfy omnigenity (and its stricter subclass, quasisymmetry), with correspondingly improved neoclassical confinement \cite{sanchez_2023, goodman_constructing_2023,goodman_quasi-isodynamic_2024,dudtMagneticFieldsGeneral2024,landreman_magnetic_2022,liu_optimizing_2025}.

In the collisionless limit, neoclassical losses are dominated by trapped-particle drifts. 
In an \emph{omnigenous} magnetic field, the orbit-averaged radial drift vanishes for every trapped trajectory \cite{cary_helical_1997}, which can be enforced through stringent geometric constraints on $|B|$ in Boozer coordinates; near-omnigenous (and quasisymmetric) optimization has therefore achieved substantially improved confinement \cite{sanchez_2023, goodman_constructing_2023,goodman_quasi-isodynamic_2024,dudtMagneticFieldsGeneral2024,landreman_magnetic_2022,liu_optimizing_2025}.

The bootstrap current is a net current parallel to the magnetic field that is self-generated within the plasma, without any external drive, due to toroidicity. 
The presence of any plasma current can alter the magnetic equilibrium, a typical effect being the modification of the magnetic island divertor structure. 
Poloidal omnigenous (PO) stellarators (also termed ``quasi-isodynamic'' stellarators) are a type of omnigenous stellarator in which the contours of constant magnetic field strength close in the poloidal direction. 
As a consequence of this, they are characterized by a zero bootstrap current at low collisionality \cite{helander_bootstrap_2009}.
Therefore, this concept has garnered considerable attention. In particular, it is the foundation for the world's largest operating stellarator, Wendelstein 7-X, and is a cornerstone of most stellarator reactor concepts \cite{sanchez_2023, sanchez2025ciematqi4xreactorrelevantquasiisodynamicstellarator,goodman_quasi-isodynamic_2024, LION2025114868, Hegna_Anderson_Andrew_Ayilaran_Bader_Bohm_Mata_Canik_Carbajal_Cerfon_et}. 
However, the requirement that $B$-contours be poloidally closed is arguably at odds with the toroidal nature of the configuration. 
Consequently, optimized PO configurations often exhibit either a very large mirror ratio or a significant elongation in the high-field region\cite{rodriguez_near-axis_2025}, accompanied by complex coil designs and an economically unfavorable large aspect ratio. 
The design of a stellarator reactor with a PO configuration necessitates a series of challenging trade-offs across multiple performance metrics\cite{cadena_constellaration_2025}.

% pwO
By relaxing some of the requirements that omnigenity imposes on the spatial variation of $B$, a novel family of stellarator magnetic fields, known as piecewise omnigenous (pwO) fields, has been recently proposed \cite{velasco_piecewise_2024}. 
These magnetic fields can effectively show reduced neoclassical transport and achieve good particle confinement even when the magnetic field distribution do not have $B$-contours that close poloidally, helically or toroidally \cite{bindel_2023}. 
Interestingly, recent analytical work, confirmed by numerical simulations for model fields, has proven the existence of piecewise omnigenous fields that yield zero bootstrap current at low collisionality for any density and temperature profiles\cite{calvo_piecewise_2025}, being therefore in principle compatible with an island divertor. 

The pwO concept may thus radically expand the range of configurations that can be candidates for fusion reactors, some of which might be easier to design or build. 
Nevertheless, as pwO is a nascent concept, optimization of this class of magnetic fields is still in its early stages, and comprehensive research is required to bring it to maturity.  
Only a few configurations have so far been reported as exhibiting features associated with pwO, including 
1) configurations A and B \cite{bindel_2023}, obtained through direct optimization of fast ion orbits; 
2) the US131 vacuum umbilic stellarator\cite{gaur_omnigenous_2025}, obtained through direct optimization of the effective ripple;
3) the OOPS-PO-pwO configurations\cite{liu_optimizing_2025}, which approximate pwO on the high-field side while the low-field part remains PO; 
4) and the CIEMAT-pw1 configuration\cite{fernandez_pacheco_albandea_piecewise_nodate}, derived by minimizing the deviation between the equilibrium magnetic field and a \textit{prototypical} pwO field. 

Relaxing high-field-side constraints is a natural way to increase geometric flexibility, but unconstrained optimization can markedly worsen confinement by disrupting the $B_\text{max}$ structure. 
In our early work \cite{liu_optimizing_2025}, we adopted a controlled relaxation.
The homeomorphic mappings preserve omnigenity while the mapping range is limited.
The $B_\textbf{max}$ regions close locally and the omnigenity mapping ensures that the fieldlines connect ``parallelogram'' corners.
This approach proves to be effective and the new PO-pwO configurations demonstrate excellent confinement.
It is naturally interpreted within the picture of \cite{velasco_exploration_2025} and \cite{Velasco2026} (there, the term QI-pwO is employed). The former paper proposed combinations of omnigenous and piecewise omnigeneous fields; the latter, more recent, investigates the PO-pwO parameter space with the constraint of zero bootstrap current.
Altogether, we have shown that there exists a region of the configuration space in which partially-relaxed omnigenity can alleviate shaping and coil complexity, while neoclassical confinement is retained.
In this paper, we further explore the PO-pwO design space and present various PO-pwO configurations with favourable confinement properties.

% Along this line, it should be noted that the concept of piecewise omnigenity also helps explain some of the transport properties of the standard configuration of Wendelstein 7-X and the inward-shifted configuration of the Large Helical Device (LHD)\cite{yamada_characterization_2005}. These two configurations display less shaping and, at the same time, lower neoclassical transport than other configurations that are closer to fulfilling the constraints imposed by omnigenity. Both configurations have sometimes been termed 'approximately omnigenous' or 'quasi-omnigenous' (QO) \cite{mynick}. The models of \cite{velasco_exploration_2025} provide a rigorous framework in which approximately omnigenous or QO fields can be made compatible with a reactor scenario, and the work presented in this manuscript will be its first practical implementation.

The remainder of this paper is organized as follows. 
Section~\ref{sec:method} describes the \texttt{OOPS} framework and the omnigenity mapping used to construct PO–pwO configurations via the ``squeeze’’ procedure. 
% we will first review an efficient \texttt{OOPS} optimization method and its application in omnigenity optimization, and subsequently extend it to the construction of nearly-PO pwO configurations (from now on, PO-pwO). 
Section~\ref{sec:results} presents representative optimized PO–pwO configurations that demonstrate their potential for fusion reactor applications. 
Section~\ref{sec:summary} summarizes the main conclusions and outlines directions for future work.

\section{Method}\label{sec:method}
\subsection{The \texttt{OOPS} method}
% Inspired by the success of QS optimization, we introduce a new approach for omnigenity optimization \texttt{OOPS} (Omnigenity OPtimization like quasiSymmetry)\cite{liu_optimizing_2025}. For magnetic fields with \textit{no} locally closed $B$ contours, such as QS, omnigenity, and pseudosymmetry, one can define a homeomorphic coordinate system $(\alpha, \eta)$ in which the $B$ contours become straight along the $\alpha$ direction. The homeomorphism between $(\alpha, \eta)$ and $(\theta_B, \zeta_B)$ is yet undetermined and differs for QS, omnigenity, and other configurations.
% In the case of QS, the homeomorphism is relatively straightforward, since $B$ contours are straight in Boozer or Hamada coordinates. A simple choice is $\theta_B = \alpha, \zeta_B = \eta$ .

Inspired by the success of quasisymmetry optimization, we introduce \texttt{OOPS} (Omnigenity OPtimization like quasiSymmetry) as a general approach to optimizing omnigenity and other concepts \cite{liu_optimizing_2025}. 
For magnetic fields that do not have locally closed $B$-level sets—such as quasisymmetric, omnigenous, and pseudosymmetric fields—there exists a homeomorphic change of coordinates $(\theta_B,\zeta_B)\mapsto(\alpha,\eta)$ that straightens the $B$-contours so they run parallel to the $\alpha$ direction. 
The particular homeomorphism is not fixed a priori and depends on the configuration class (quasisymmetry, omnigenity, etc.). 
In the quasisymmetric case, the mapping is particularly simple: because $B$-contours are already straight in Boozer or Hamada coordinates, one may take $(\alpha,\eta)=(\theta_B,\zeta_B)$ .

% For omnigenity, the homeomorphism must be carefully constructed to ensure the required omnigenity conditions are satisfied.
% Previous works by Cary \& Shasharina \cite{cary_helical_1997}, Landreman \& Catto \cite{landremanOmnigenityGeneralizedQuasisymmetrya2012}, and Dudt \etal \cite{dudtMagneticFieldsGeneral2024} offer different strategies to construct ideal omnigenous fields.
% They can all be understood as different homeomorphisms for omnigenity.

For omnigenity, the homeomorphic change of coordinates must be chosen so that the omnigenity constraints are enforced. 
Classical constructions by Cary \& Shasharina \cite{cary_helical_1997}, Landreman \& Catto \cite{landremanOmnigenityGeneralizedQuasisymmetrya2012}, Parra \etal \cite{parra_less_2015}, and the more recent framework of Dudt \etal \cite{dudtMagneticFieldsGeneral2024} provide complementary recipes for realizing ideal omnigenous fields.
\texttt{OOPS} has also proposed an omnigenity mapping.
Viewed through this lens, these works correspond to distinct realizations of the omnigenity-straightening homeomorphism.

Once the coordinate mapping from $(\alpha,\eta)$ to $(\theta_B, \zeta_B)$ has been established, omnigenity can be realized by minimizing the asymmetric modes,
\begin{equation} \label{eq:fomni}
    f_\text{omni} = \sum_{m \neq 0} (B_{m,n}/B_{0,0})^2 \ ,
\end{equation}
where $B(\alpha,\eta) = \sum B_{m,n} \cos(m\alpha - n\eta)$ and $B_{m,n}$ are computed with Fourier decomposition.
Compared to pointwise matching methods, \texttt{OOPS} only enforces the symmetry direction and thus can explore more parameter space.

\subsection{The Landreman-Catto omnigenity mapping}
Any of the omnigenity-straightening homeomorphisms discussed above can be used within the \texttt{OOPS} framework. 
For the PO-pwO configurations, we find it easier to use the Landreman–Catto  mapping\cite{landremanOmnigenityGeneralizedQuasisymmetrya2012}.
We have intermediate coordinates, $(\tilde{\theta},\tilde{\zeta})$, which will be explained later. 
The Landreman-Catto omnigenity mapping is defined as $\tilde{\theta} = \alpha$ and
\begin{equation}\label{eq:LC-mapping}
  \tilde{\zeta}(\eta, \alpha)= 
    \begin{cases}
        \pi-s(\eta, \tilde{\theta}+\tilde{\iota} D(\eta))-D(\eta) 
        & \text { if } 0 \leq \eta \leq \pi \\ 
        \pi+s(2 \pi-\eta,-\tilde{\theta}+\tilde{\iota} D(2 \pi-\eta))+D(2 \pi-\eta) 
        & \text { if } \pi<\eta \leq 2 \pi
    \end{cases}
\end{equation}
where the function $s(x,y)$ represents the $\zeta$-variation of the $B$ contours and $D(\eta)$ determines the bounce distance.
$s$ has a periodicity of $2\pi$ in $y$ and is an odd function of $y$. The input $x$ ranges over $[0,\pi]$ and require $s(0,y)=0$ for all $y$. 
The function $D(x)$ defined on $[0,\pi]$ and satisfy $D(0) = \pi$ and $D(\pi)=0$. 
Here, the quantities $(\tilde{\theta},\tilde{\zeta},\tilde{\iota})$ are called effective quantities. 
Additional transformations are used to transform these back to the original Boozer coordinates in toroidal, poloidal, and helical directions. 
For poloidally and helically omnigenous fields, the effective quantities are defined as $\tilde{\theta} = \theta$, $\tilde{\zeta} = (N\zeta-M\theta)N_\text{fp}$, and $\tilde{\iota} = \iota/[(N-\iota M)N_\text{fp}]$ ($M, N$ determine the helicity). 
For toroidally omnigenous fields ($(M,N)=(1,0)$) we use $\tilde{\theta} = N_\text{fp}\zeta$, and $\tilde{\iota} = N_\text{fp}/\iota$.

% Landreman-Catto Mapping has a very straight forward view for constructing a target omnigenous field. This advantage depends de-coupled between the Shape funcion $s(x,y)$ and Distance function $D(\eta)$. In an perfect omnigenous field, bounce distance is the function of $(\alpha,B^*)$, expressed in this mapping, the separations in $\zeta$ between the pair of points on opposite branches of a field line but at the same $B$ satisfied $\Delta_\zeta = 2D(\eta)$. Thus simpliy change 

Within the Landreman-Catto mapping, an omnigenous field can be defined in a particularly flexible way because the construction decouples the shape function $s$ and the distance function $D$. 
The function $s$ controls how a given B-level set is embedded in the $(\tilde{\theta},\tilde{\zeta})$ plane, while $D(\eta)$ fixes the spacing between the two opposite branches of a field line at the same value of $B$. 
In an exactly omnigenous field generated by the Landreman-Catto mapping, the toroidal separation between those branch points depends only on $\eta$ (equivalently, only on $B$), namely
$\Delta\zeta(\eta) = 2D(\eta)$. 
Hence, by prescribing any admissible pair of $s$ and $D$, one can synthesize families of fields whose $B-$contours have the desired geometry and satisfy omnigenity.

\subsection{Parameterization of pwO fields}
The variation of the magnetic-field strength for a prototypical piecewise-omnigenous (pwO) field~\cite{velasco_piecewise_2024} can be written as
\begin{equation}
\begin{aligned}
B_{pwO}
&= B_{\min} + \left(B_{\max}-B_{\min}\right)\times\\
&\lim_{p\to\infty}
\exp\!\left[
-\left(\frac{\zeta_B-\zeta_c+t_1(\theta_B-\theta_c)}{w_1}\right)^{2p}
-\left(\frac{\theta_B-\theta_c+t_2(\zeta_B-\zeta_c)}{w_2}\right)^{2p}
\right], \\
\iota
&=
\left(
\frac{\pi(1-t_1 t_2)}{N_{\mathrm{fp}} w_1}-1
\right)^{-1} t_2 .
\end{aligned}
\end{equation}
In this representation, contours of constant \(B\) are parameterized by two families of parallel lines. 
One family has slope \(-1/t_1\) and is toroidally separated by a distance \(2w_1\), while the other has slope \(-t_2\) and is poloidally separated by \(2w_2\). 
The center of the corresponding parallelogram is located at \((\zeta_c,\theta_c)\).
The parameter \(p\) controls how closely the high-field region approaches an ideal parallelogram. 
In the formal limit \(p\to\infty\), the boundary becomes sharp, recovering the idealized pwO construction. 
However, in practice, taking \(p\) to be sufficiently large is already enough to reproduce the essential physical properties of pwO fields while slightly relaxing the discontinuous structure, which is advantageous for numerical optimization and field construction \cite{escoto_evaluation_2025}.

During the optimization, the equilibrium's rotational transform \(\iota_0\) is prescribed. 
Therefore, $w_1$ is no longer treated as an independent parameter, but can instead be determined by $(t_1,t_2,\iota_0)$ through
\begin{equation}
w_1(t_1,t_2,\iota_0)=
\frac{\pi}{N_{\mathrm{fp}}}
\frac{1-t_1 t_2}{1+t_2/\iota_0}.
\end{equation}\label{EQ_W1}
(note that other relations are possible \cite{velasco_exploration_2025}). In addition, it has been shown that a pwO field without bootstrap current must satisfy \cite{calvo_piecewise_2025}
\begin{equation}
w_2 = \pi.
\end{equation}
if $w_1$ is set by equation (\ref{EQ_W1}). Accordingly, once the equilibrium's $\iota_0$  and the zero-bootstrap-current condition are imposed, the pwO geometry is fully specified by the two parameters \((t_1,t_2)\), with \(w_1\) determined from the relation above and \(w_2\) fixed to \(\pi\).

\subsection{Generation of PO-pwO}
% Squeeze process
The pwO concept arises from $\partial \mathcal{J} / \partial\alpha = 0$ being fulfilled piecewisely \cite{velasco_piecewise_2024}, being $\mathcal{J}$ the second adiabatic invariant. This explains good confinement in configurations deviating from exact omnigenity. 
However, most pwO fields explored so far \cite{velasco_piecewise_2024,calvo_piecewise_2025,fernandez_pacheco_albandea_piecewise_nodate} feature locally-closed $B_\text{max}$ contours and flattened, single-valued $B_\text{min}$ regions.
This implies large gradients in the field strength at the transition parts between $B_\text{min}$ and $B_\text{max}$.
At low aspect ratios, realizing such a structure may be particularly challenging as strong toroidicity makes it difficult to form uniform $B$.
A PO-pwO stellarator magnetic field \cite{velasco_exploration_2025,liu_optimizing_2025,Velasco2026} allows smooth variations in $B$ from $B_\text{min}$ region to the $B_\text{max}$ region, with $J$ being constant on the flux surface for deeply trapped particles and piecewisely constant for the rest.
Here, we show how to numerically optimize PO-pwO configurations.

% To address this, we consider pwO fields closer to omnigenity \cite{velasco_exploration_2025}: specifically, configurations that behaves as quasi-isodynamic (in particular, with $B_{min}$- contours closing poloidally) for deeply trapped particles while retaining pwO behavior for the remaining orbits.
% These fields can be viewed as a controlled relaxation of omnigenity and, importantly, remain compatible with the homeomorphic straightening framework within their omnigenous subdomains.
% Given the requirements for steady-state operation, especially elimination of bootstrap currents, PO is our primary design objective for deeply trapped particles (although it should be noted that control of the pwO part of the field is required for cancellation of the bootstrap at low collisionality \cite{calvo_piecewise_2025, Velasco2026}. Operationally, the construction reduces to controlling the straightening map across the interfaces between the PO and pwO regions (the “squeezed” interfaces). 
% Since the $B_{\min}$ regions remain omnigenous and B varies smoothly, the corresponding $B_{\max}$ regions satisfy the pwO condition on rotational transform($\iota$), as we will see.

We develop this idea within the Landreman–Catto mapping (Eq.~\ref{eq:LC-mapping}) by restricting the range of $\eta$ and modifying the distance function profile $D$.
% For pwO, we choose omnigenity mapping functions with strongly shaped $B$ contours and limit the range of $\eta$ so that $B_{\text{max}}$ contours will be ``squeezed'' to form locally closed ``parallelograms''.
For pwO, we start from omnigenous mapping functions that produce strongly shaped $B$-contours, and then limit the accessible $\eta$-range so that the $B_{\text{max}}$ contours are “squeeze”, forming locally-closed, parallelogram-like structures. 
Because the $B_{\text{min}}$ regions remain omnigenous and the magnetic field varies smoothly across the space, the resulting $\iota$ satisfies the pwO condition.

\begin{figure}
    \centering
    \includegraphics[width=0.8\textwidth]{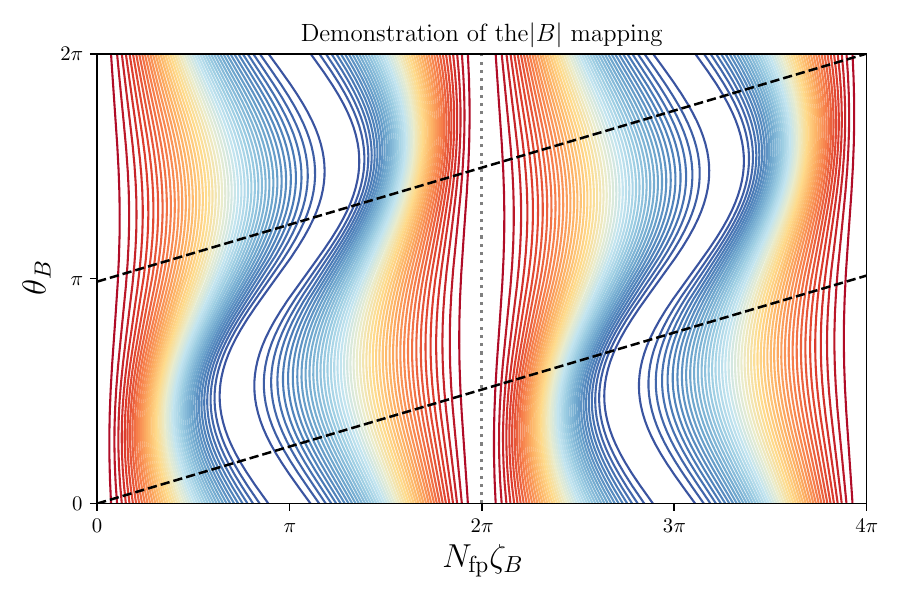}
    \caption{The field strength $B$ of a poloidal-omnigenous magnetic field in Boozer coordinates; the black dashed line shows a straight-field-line trajectory.\label{fig:PO}}
\end{figure}

%We now illustrate how this squeezing operation is implemented in the Landreman–Catto mapping. 
As a representative PO setup, we consider a configuration with $N_\text{fp}=3$ and $\iota = 0.76$, and we choose
\begin{equation}
    S(x,y) = 0.3 x \sin y, \quad D(x) = \pi - x ,
\end{equation}
while in $(\eta,\alpha)$ coordinates, we prescribe an arbitrarily PO magnetic field of the form
\begin{equation}
    B(\eta,\alpha) = 1 + 0.25\cos\eta \ .
\end{equation}
The mapping then yields the distribution of $B(\theta_B,\zeta_B)$ in Boozer coordinates, which is shown in Fig.~\ref{fig:PO}.

\begin{figure*}
    \centering
    \includegraphics[width=1\textwidth]{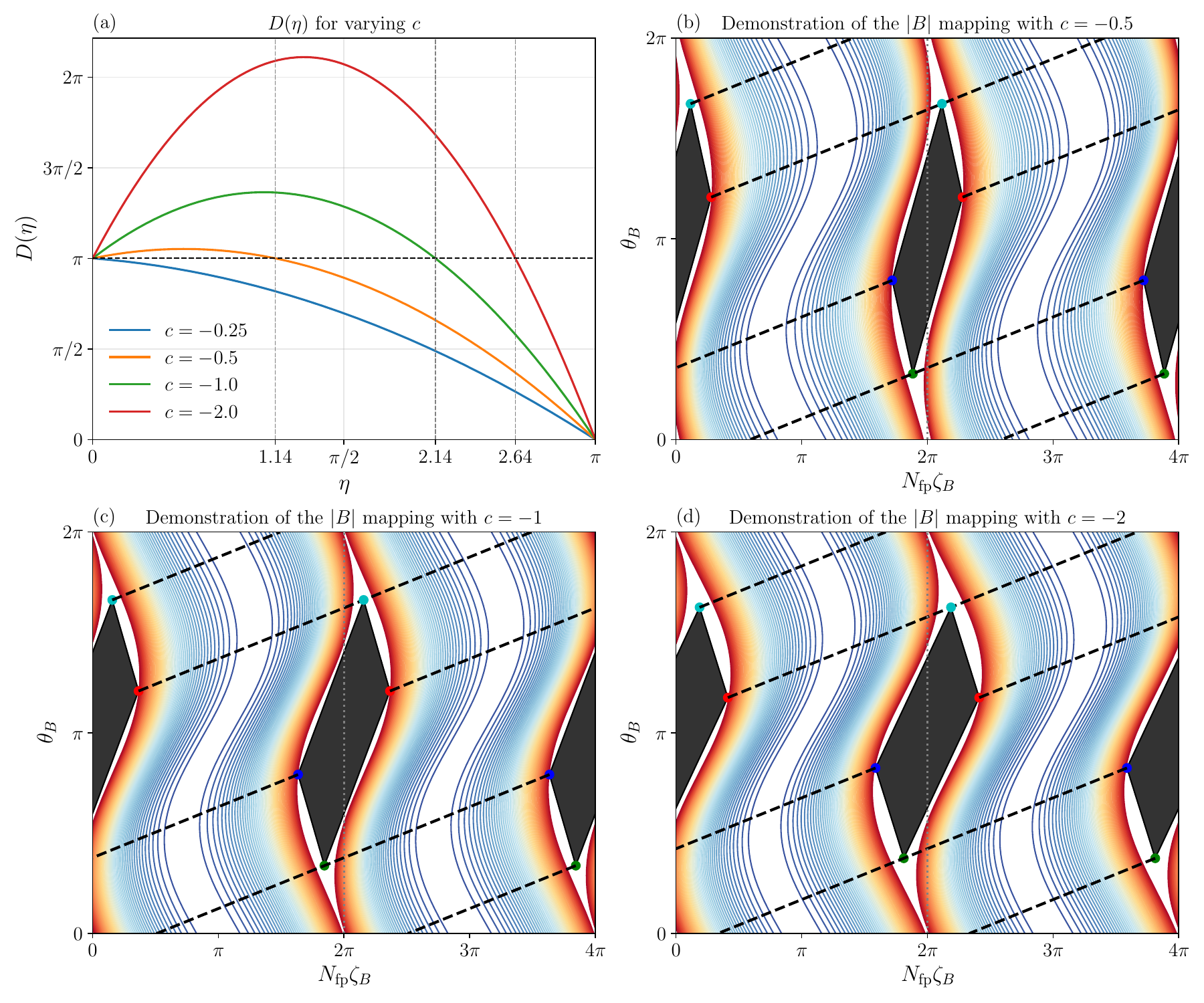}
    \caption{Effect of varying D on the ``squeeze'' process with all other mapping parameters held fixed. The region enclosed by the outermost contour is manually fitted using a zero-bootstrap-current pwO mapping ($w_2=\pi$) in \cite{calvo_piecewise_2025}. (a) Changing the parameter $c$ deforms $D(x)$, thereby altering the degree of compression of the bounce-angle span. (b) For $c=-0.5$, the PO-mapping coordinate is compressed by $\Delta\eta=1.65$; a plausible pwO region within the unmapped interval is characterized by $(t_1,t_2)=(-0.15,\,8.5)$. (c) For $c=-1$, the compression is $\Delta\eta=2.34$; a candidate pwO region is $(t_1,t_2)=(-0.2,\,6.5)$. (d) For $c=-2$, the compression is $\Delta\eta=2.74$; a candidate pwO region is ($t_1,t_2)=(-0.25,\,6)$.\label{fig:Dfunc}}
\end{figure*}

In the construction, the geometry of equal-bounce-angle contours associated with omnigenity is governed by the function $D$ via
\begin{equation}
    \Delta\zeta = 2D(\eta) \ .
\end{equation}
The essence of the ``squeeze'' is to modify $D$ such that the full $2\pi$ variation in $\Delta\zeta$ occurs before $\eta$ reaches $0$ or $2\pi$. 
Since the function $S$ has not yet been relaxed to zero at that point, the mapped image contains an isocontour that closes on itself, realizing the omnigenous condition $\partial{\mathcal{J}}/\partial\alpha = 0$.
As a concrete example, one can use
\begin{equation}
    D(\eta) = c\eta^2 - (1+c\pi)\eta + \pi \ .
\end{equation}
In this form, if $\eta=x_{\mathrm{point}}=(1+c\pi)/c$, we have $D= \pi$ and $\Delta \zeta = 2 \pi$, as shown in Fig~\ref{fig:Dfunc}(a).
When we restrict $\eta\in[x_{\mathrm{point}},\pi]$ and use the omnigenity mapping, the resulting ``squeezed"-omnigenous field $B(\theta_B,\zeta_B)$ has an outermost isocontour that closes locally.
Furthermore, along this contour, the equal-$\mathcal{J}$ (constant bounce action) condition is still satisfied. 
The region enclosed by this contour has the shape of the pwO fields discussed in \cite{velasco_piecewise_2024}.
Although no direct optimization is performed inside that region, the construction demonstrates compatibility with pwO: one can draw a parallelogram whose corners are connected to field-lines. 
The parameter $c$ can vary to form parallelograms in different shapes.
Fig.~\ref{fig:Dfunc}(b-d) illustrates several mapping variants obtained by tuning the parameters of different $D$-functions, together with the resulting closed contours. The black shaded region indicates the approximate domain generated under the pwO mapping; detailed definitions of the corresponding equations and parameters can be found in Ref.~\cite{velasco_exploration_2025}

\begin{figure}
    \centering
    \includegraphics[width=1\linewidth]{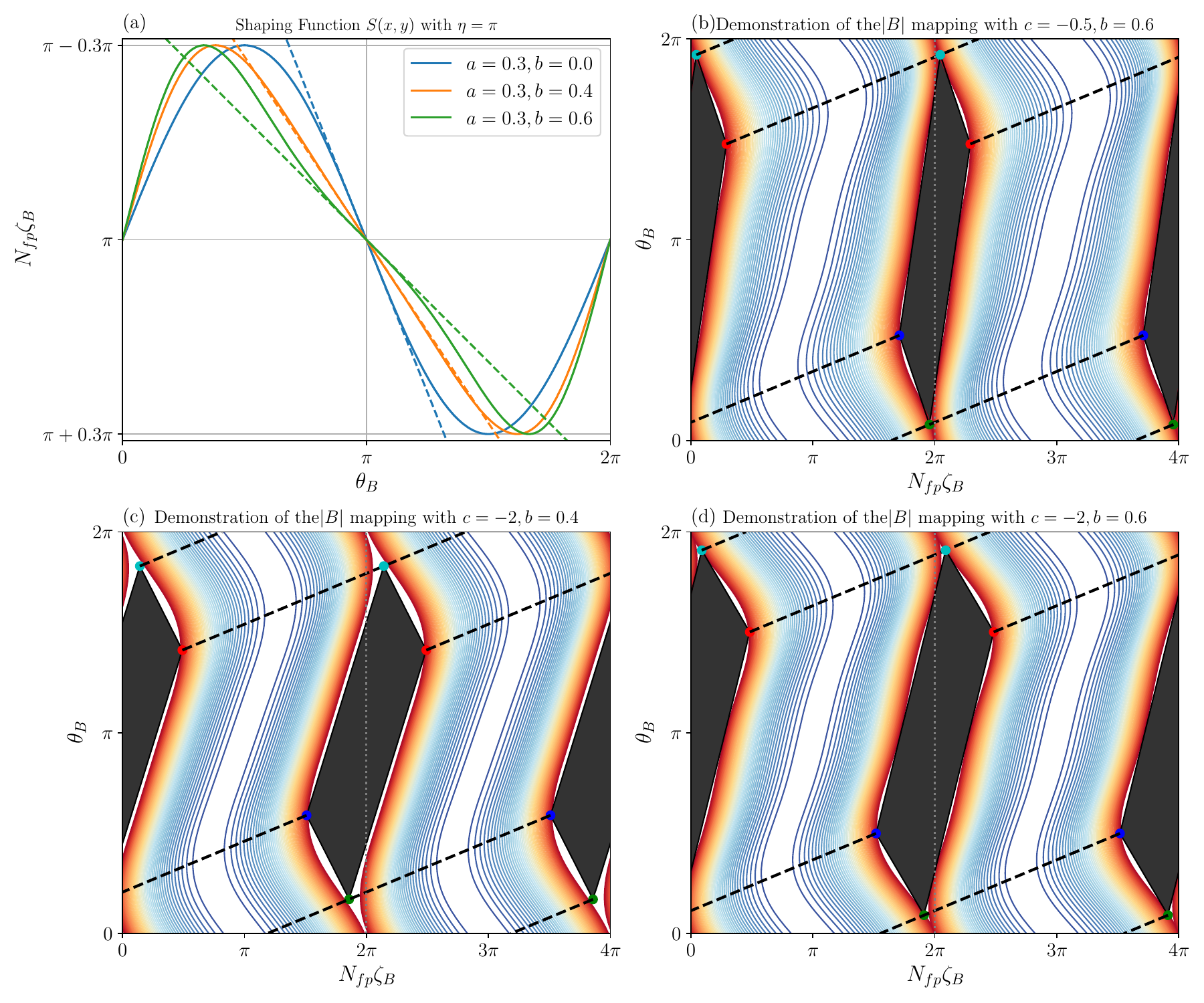}
    \caption{Effect of varying $S$ on the ``squeeze" process. For all examples,  $a=0.3$ is fixed. The region enclosed by the outermost contour is manually fitted using a zero-bootstrap-current pwO mapping ($w_2=\pi$). (a) Dependence of the apparent straightness on $b$. The colored dashed line shows the tangent at the point of interest; the degree of overlap serves as a visual proxy for local linearity. (b) For $c=-0.5, b=0.6$, the PO-mapping coordinate is compressed by $\Delta\eta=1.65$; a plausible pwO region within the unmapped interval is characterized by $(t_1,t_2)=(-0.08,\,5.4)$. (c) For $c=-2, b=0.4$, the PO-mapping coordinate is compressed by $\Delta\eta=2.76$; a plausible pwO region within the unmapped interval is characterized by $(t_1,t_2)=(-0.17,\,3.6)$. (d) For $c=-2, b=0.6$, the PO-mapping coordinate is compressed by $\Delta\eta=2.76$; a plausible pwO region within the unmapped interval is characterized by $(t_1,t_2)=(-0.135,\,3.1)$.
    \label{fig:S_function}}
\end{figure}

The presence of a vacuum magnetic well is widely expected to enhance the ideal MHD stability\cite{greene_brief_1998}. 
In the \texttt{OOPS} study, we observed that certain choices of the mapping help drive the optimizer to spontaneously develop a vacuum magnetic well, while preserving excellent omnigenity. 
In particular, there exists a family of $S$-functions that, when included as an objective term, consistently yield both good omnigenity and a vacuum magnetic well:
\begin{equation}
    S(x,y) = a x\sin(y+b\sin(y)) \ .
\end{equation}
The nested-sine form, inspired by the LHD coil parameterization \cite{okamuramagnetic}, embeds quasi-linear segments within an otherwise smooth periodic waveform.
As a result, the mapped geometry near the boundary more closely approximates the parallelogram construction characteristic of ideal piecewise omnigenity (pwO), as shown in Fig.~\ref{fig:S_function}.
% Notably, a poloidal-omnigenous (PO) mapping can be generated directly from an ideal pwO mapping\cite{velasco_exploration_2025}.

Accordingly, by constraining the functional form of the conventional omnigenity mapping and restricting the domain of the coordinate transformation, we ensure that the omnigenity constraints are fully preserved, while simultaneously creating a region that approximates/is a piecewise-omnigenous (pwO) region. 
The conjunction of these two features promotes the emergence of pwO during optimization. 
Although the resulting structures are not always as idealized as those illustrated above, the computed equilibria exhibit good neoclassical transport properties (it has been shown in \cite{fernandez_pacheco_albandea_piecewise_nodate} that local deviations of 10\% from an exact pwO field are compatible with reactor-grade confinement). Detailed evidence is reported below.

% In pwO optimization, Eq.~\ref{eq:OOPS-mapping} is revised to
% \begin{equation}
%     \tilde{\zeta} = \eta^* - (\pi -|\eta|)s_1 \sin[y(\alpha, \eta^*)] \  ,
% \end{equation}
% where $\eta^* = \eta(c|\eta|-c\pi+1)$ with $c$ as a numerical factor. 

% explanation (Velasco's figure)

\section{Numerical results}\label{sec:results}
In this section, we present a series of PO–pwO configurations constructed using the method described above, and evaluate various transport-related physical quantities to illustrate the performance of the optimization scheme in detail. 
%We begin by revisiting the PO–pwO configuration introduced in our previous work. 
We provide an initial survey of PO–pwO configurations with favorable properties over a range of field periods and aspect ratios, highlighting the generality of the approach. 
We also demonstrate a case in which the mapping is modified to obtain a vacuum magnetic well.

Given a prescribed stellarator plasma boundary together with specified pressure and current profiles, the corresponding equilibrium magnetic field can be computed directly \cite{kruskal_equilibrium_1958}. 
For fixed pressure and current profiles, the plasma boundary shape completely determines the magnetic-field structure. 
In practice, the boundary shape is typically represented by Fourier coefficients (with stellarator symmetry enforced):
\begin{equation}
    \begin{aligned}
        R(\theta,\phi) = \sum_{m,n}{R}_{m,n}\cos(m\theta-N_{\text{fp}}n\phi) , \\
        Z(\theta,\phi) = \sum_{m,n}{Z}_{m,n}\sin(m\theta-N_{\text{fp}}n\phi) . \\
    \end{aligned}
\end{equation}

Therefore, during the optimization, the free parameters are the boundary Fourier coefficients ${(R_{m,n}, Z_{m,n})}$. 
Using the \texttt{SIMSOPT} optimization framework\cite{landreman_simsopt_2021} with the equilibrium computed by the \texttt{VMEC} code\cite{hirshman_steepestdescent_1983}, we employ the so-called resolution-increase strategy, in which the resolution of the boundary Fourier spectrum is gradually increased.

\subsection{Diversity of PO-pwO Configurations} % some data base
\label{sec:moreconfig}
% 3x3 boozer plots
% 3x3 boundary plots
% LgradB plots -- not necessary not for optimized
% particle loss and EffitivateRipple
% partial J partial alpha

A representative PO-pwO configuration discussed in our previous work\cite{liu_optimizing_2025}, with an aspect ratio $A_p = 6$ and $N_{\text{fp}} = 3$ field periods, was optimized using the ``squeeze'' approach within the \texttt{OOPS} mapping.
We aimed to optimize this configuration by targeting parameters $a = 0.4$, $c = -0.6$, and a fixed edge rotational transform $\iota_{\text{edge}} = 0.62$.

To further illustrate the generality of the ``squeeze'' approach, the configurations presented in this section share nearly identical mapping parameters, but differ in the boundary rotational transform $\iota_\text{edge}$, the number of field periods $N_\text{fp}$, and the aspect ratio $A_p$. 
The aspect ratios $A_p$ are chosen be $2N_\text{fp} - 2 $, $ 2N_\text{fp}$, and $2N_\text{fp} + 2 $. 
For configurations with $N_\text{fp} = 2$, we only consider $A_p = 4$ and $A_p = 6$ to avoid difficulties associated with very small aspect ratios, such as flux-surface self-intersections. 
For the $N_\text{fp} = 5$ configuration, we adopt a commonly used aspect ratio $A_p = 10$. 

%We uniformly employ the Landreman–Catto mapping to optimize for PO-pwO properties. 
The mapping parameter is set to $a = 0.3$ for the $N_\text{fp} = 2, 3, 4$ configurations, and to a = 0.4 for the $N_\text{fp} = 5$ configuration. 
All other mapping parameters are kept fixed at $b = 0$, $c = -2$, and $\Delta\eta = 2.8$.
% Empirically-by means of observation or experience rather than theory or pure logic.
The target edge rotational transform $\iota_\text{edge}$ is chosen to avoid low-order rational surfaces: for the $N_\text{fp} = 2, 3, 4\text{, and }5$ configurations, we avoid the $(m,n) = (4,2), (4,3), (5,4)\text{, and } (5,5)$ rationals, respectively. 
Empirically, when the boundary is optimized solely through the mapping, without imposing additional constraints such as a vacuum magnetic well, the resulting rotational-transform profiles tend to exhibit negative shear $d\iota/d\psi<0$. 
Therefore, $\iota_\text{edge}$ is selected to lie slightly above the corresponding low-order rational values. The total objective function is given by
\begin{equation}
    f=f_{PO-pwO}+(A_p-A_p^*)^2+(\iota_{\text{edge}}-\iota_{\text{edge}}^*) \ .
\end{equation}

\begin{figure}
    \centering
    \includegraphics[width=1\linewidth]{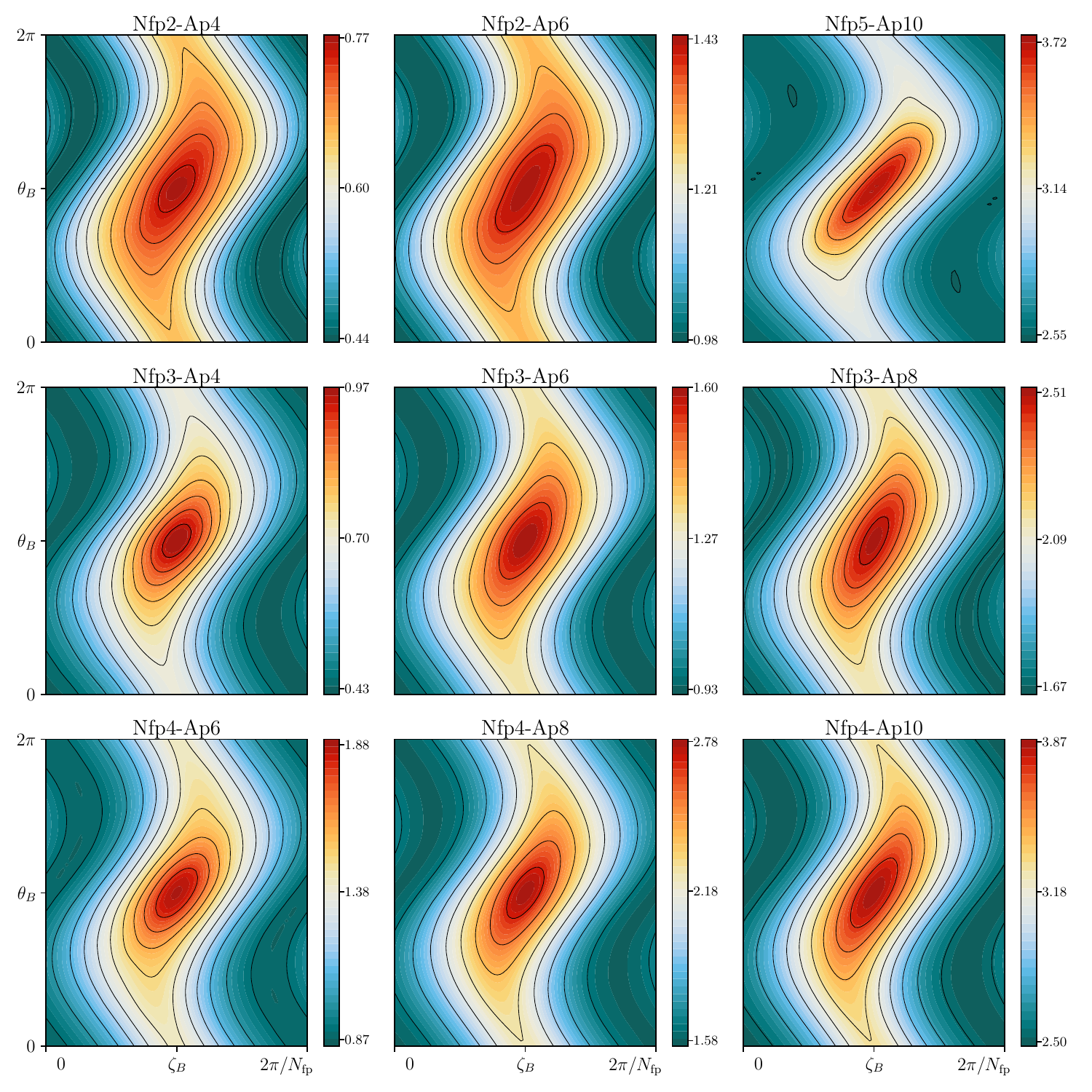}
    \caption{A gallery of PO–pwO configurations. Magnetic-field-strength distributions in Boozer coordinates on the LCFS for optimized configurations with different field periods and aspect ratios.
    \label{fig:all-boozer}}
\end{figure}

\begin{figure}[h]
    \centering
    \includegraphics[width=1\linewidth]{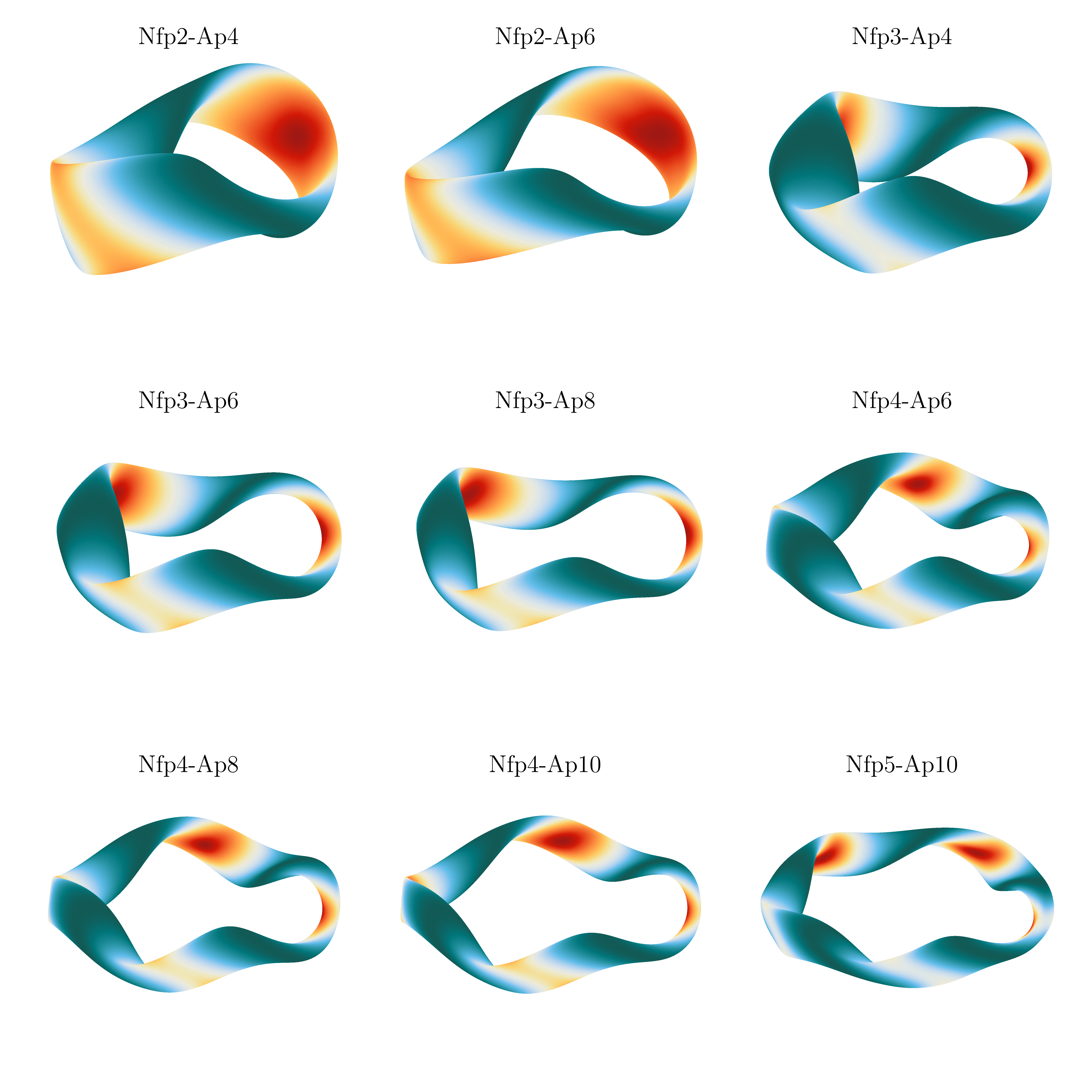}
    \caption{A gallery of PO–pwO configurations. Geometries of the LCFS for optimized configurations with different field periods and aspect ratios; colors indicate the $|\mathbf{B}|$ on the surface.
    \label{fig:all-boundary}}
\end{figure}

% spontaneously:as a result of a sudden impulse and without premeditation.
Figure~\ref{fig:all-boozer} shows the magnetic-field strength on the last closed flux surface (LCFS) in Boozer coordinates for the nine optimized configurations discussed above.
Within the region directly covered by the mapping, the configurations exhibit excellent PO properties.
In the adjacent region produced by the ``squeeze'' construction, localized closed $B_{\max}$ domains emerge spontaneously, with shapes that approximately follow the outermost closed contour of the mapped region.

\begin{figure}[t]
    \centering
    \begin{subfigure}{\linewidth}
        \centering
        \includegraphics[width=\linewidth]{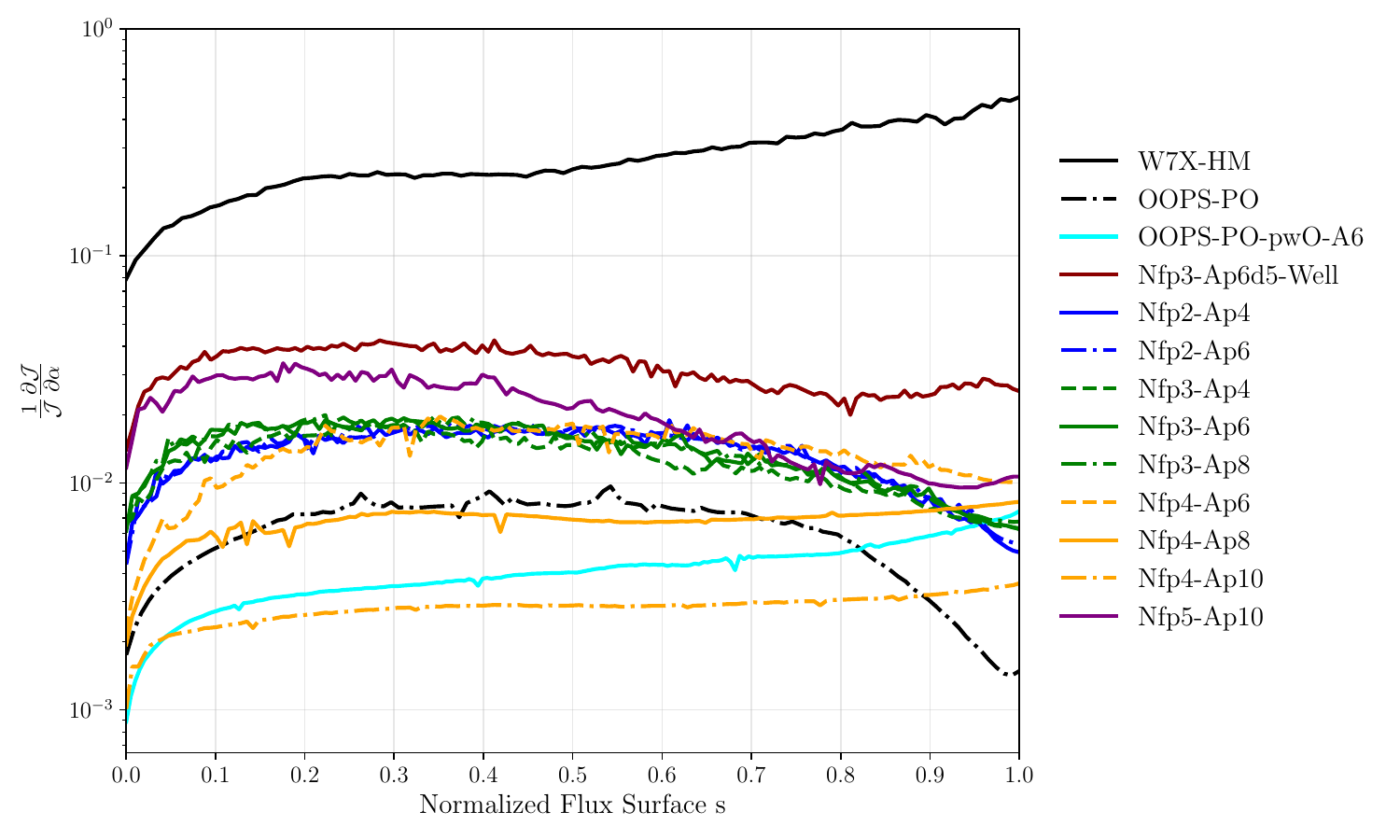}
        \caption{Departure from exact omnigenity, a measure of omnigenity arising from flux-surface geometry.}
        \label{fig:pJpa}
    \end{subfigure}
    \vspace{0.6em}
    \begin{subfigure}{\linewidth}
        \centering
        \includegraphics[width=\linewidth]{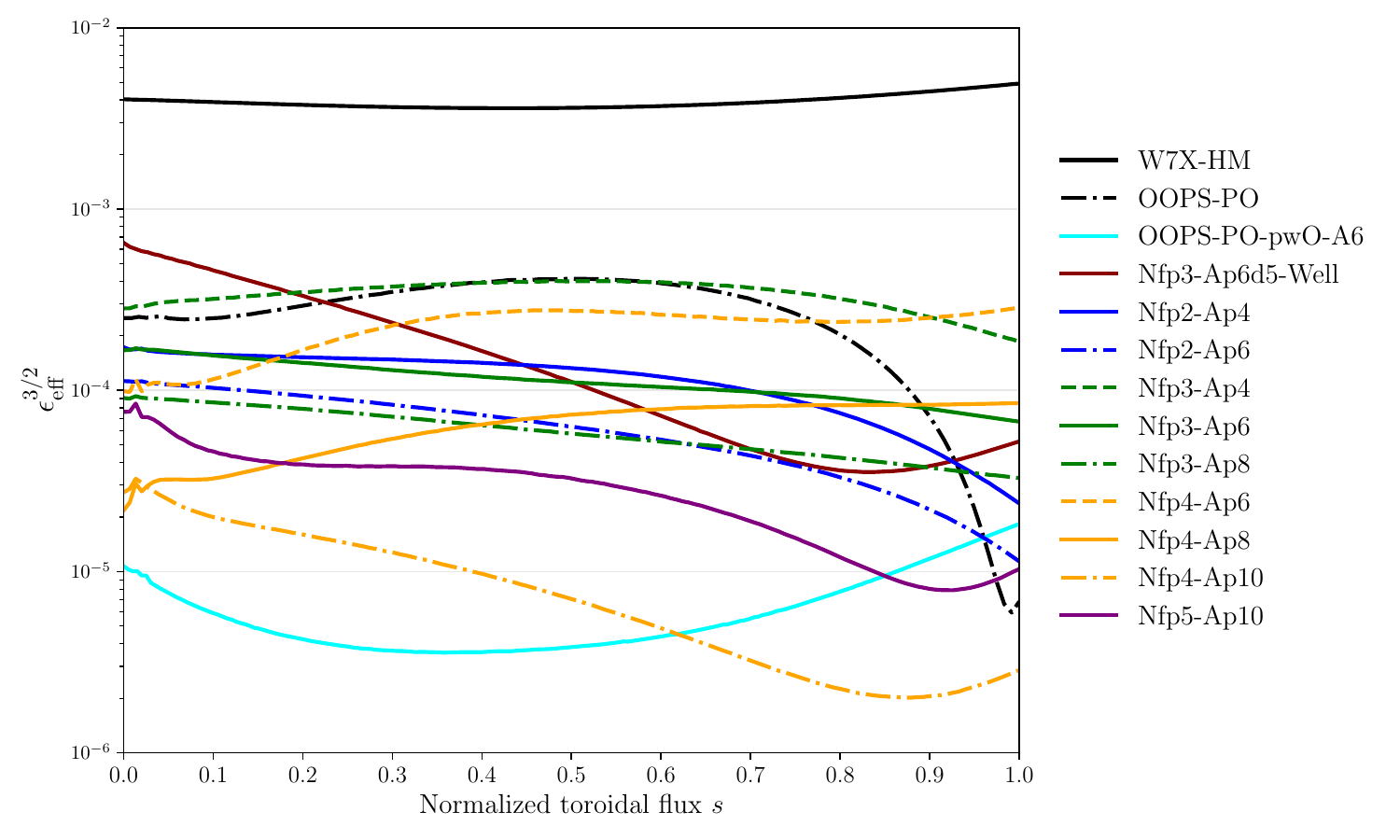}
        \caption{Effective ripple, a measure of neoclassical transport arising from flux-surface geometry.}
        \label{fig:eps}
    \end{subfigure}
    \caption{Comparison of metrics that directly quantify how close the optimized configurations are to pwO or omnigenity.}
    \label{fig:pwo_two_panels}
\end{figure}

The quality of optimized configurations is assessed by directly computing the departure from exact omnigenity\cite{boozer_required_2023}
\begin{equation}
\frac{1}{\mathcal{J}}\frac{\partial\mathcal{J}}{\partial\alpha}
=-\,\frac{
  \displaystyle \int
     \frac{\displaystyle 1-\displaystyle \frac{B}{2B^*}}{B^2\sqrt{1-\displaystyle  \frac{B}{B^*}}}
    \frac{\partial B}{\partial\alpha}\, d\zeta
}{ \displaystyle \int
    \sqrt{1-\frac{B}{B^*}}\,\frac{d\zeta}{B}
} \ ,
\label{eq:pJpa}
\end{equation}
where $B*$ is the magnetic strength at the bounce point, as well as the effective ripple $\epsilon_\text{eff}^{3/2}$\cite{nemov_evaluation_1999} for neoclassical transport.
We obtain values that are substantially lower than those of W7-X and even comparable to those of the PO configuration, as shown in Fig.~\ref{fig:pJpa} and Fig.~\ref{fig:eps}. 

\begin{figure}
    \centering
    \includegraphics[width=1\linewidth]{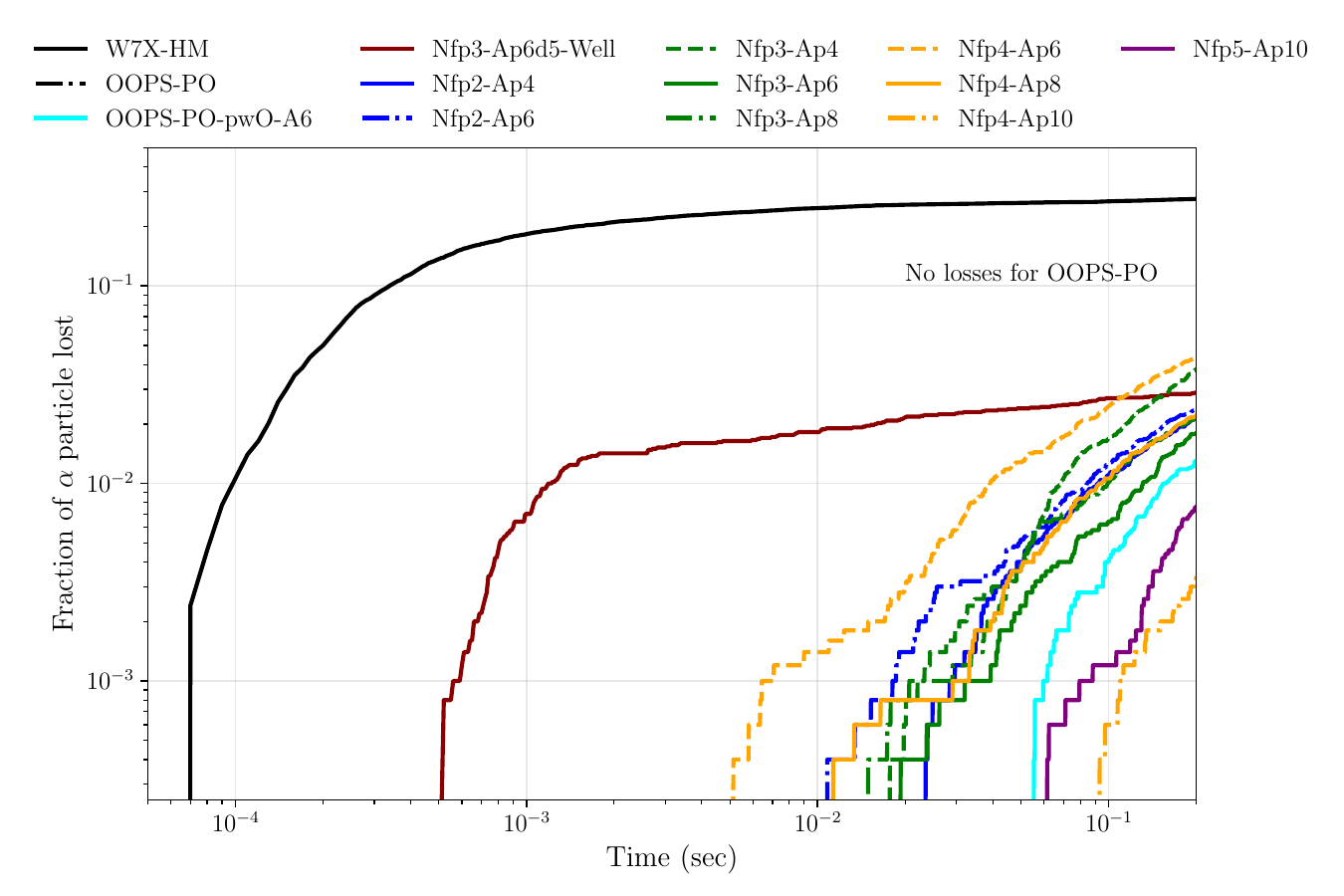}
    \caption{Loss fraction of 3.5 MeV fusion alpha particles launched from the $s=0.25$ flux surface in collisionless guiding-center simulations, scaled to reactor size.
    \label{fig:loss}}
\end{figure}

Another metric that is of interest is fast-ion confinement. 
To assess this, we scale these configurations to a minor radius of 1.7 m and a magnetic field strength of $B_0 = 5.7$ T, the same as the ARIES-CS reactor \cite{najmabadi_aries-cs_2008}, and compute the loss fraction using the \texttt{SIMPLE} code \cite{albert_symplectic_2020,albert_accelerated_2020}.
For 5000 fusion-born 3.5 MeV alpha particles launched from the $s = 0.25$ surface, the maximum alpha-particle loss over a typical slowing-down time of $0.2~\text{s}$ remains below $4\%$ for all nine cases, as shown in Fig.~\ref{fig:loss}. 
Taken together, these results demonstrate that the ``squeeze'' construction systematically produces PO-pwO configurations with consistently low effective ripple and strong energetic-particle confinement across a broad range of $N_\text{fp}$ and $A_p$. 
These results confirm that the configurations produced by the ``squeeze'' approach are genuinely a combination of omnigenity and piecewise omnigenity, rather than merely approximately omnigenous.
The \texttt{OOPS} method is a robust, broadly applicable route to generating PO-pwO stellarators rather than a finely tuned solution for a specific device.

% \begin{figure}
%     \centering
%     \includegraphics[width=1\linewidth]{figure/fig-omni_depature_comparison_pwo.pdf}
%     \caption{PO-pwO configurations depaure from omnigenous 
%     \label{fig:pJpa}}
% \end{figure}

% \begin{figure}
%     \centering
%     \includegraphics[width=1\linewidth]{figure/fig-EffectiveRipple_pwo.pdf}
%     \caption{PO-pwO configurations's Effective Ripple 
%     \label{fig:eps}}
% \end{figure}

\subsection{PO-pwO configuration with magnetic well}
We also demonstrate that PO-pwO configurations can be generated with a vacuum magnetic well solely by prescribing an appropriate mapping. 
By turning on the mapping parameter $b$, we construct the approximately parallelogram-shaped mapped region shown in Fig~\ref{fig:Dfunc}.
In this example, the mapping parameters are chosen as $a = 0.3$, $b = 0.6$, $c = -2$, and $\Delta\eta = 2.76$. 
The target edge rotational transform $\iota_\text{edge}$ is chosen in the same way as in Sec~\ref{sec:moreconfig}.
For the $N_{\text{fp}}=3$ configuration, we set $\iota_\text{edge}^* = 0.76$ and a target an aspect ratio of $A_p^* = 6.5$. 
The maximum boundary Fourier modes are kept identical to those of the previous designs.

\begin{figure}
    \centering
    \includegraphics[width=1\linewidth]{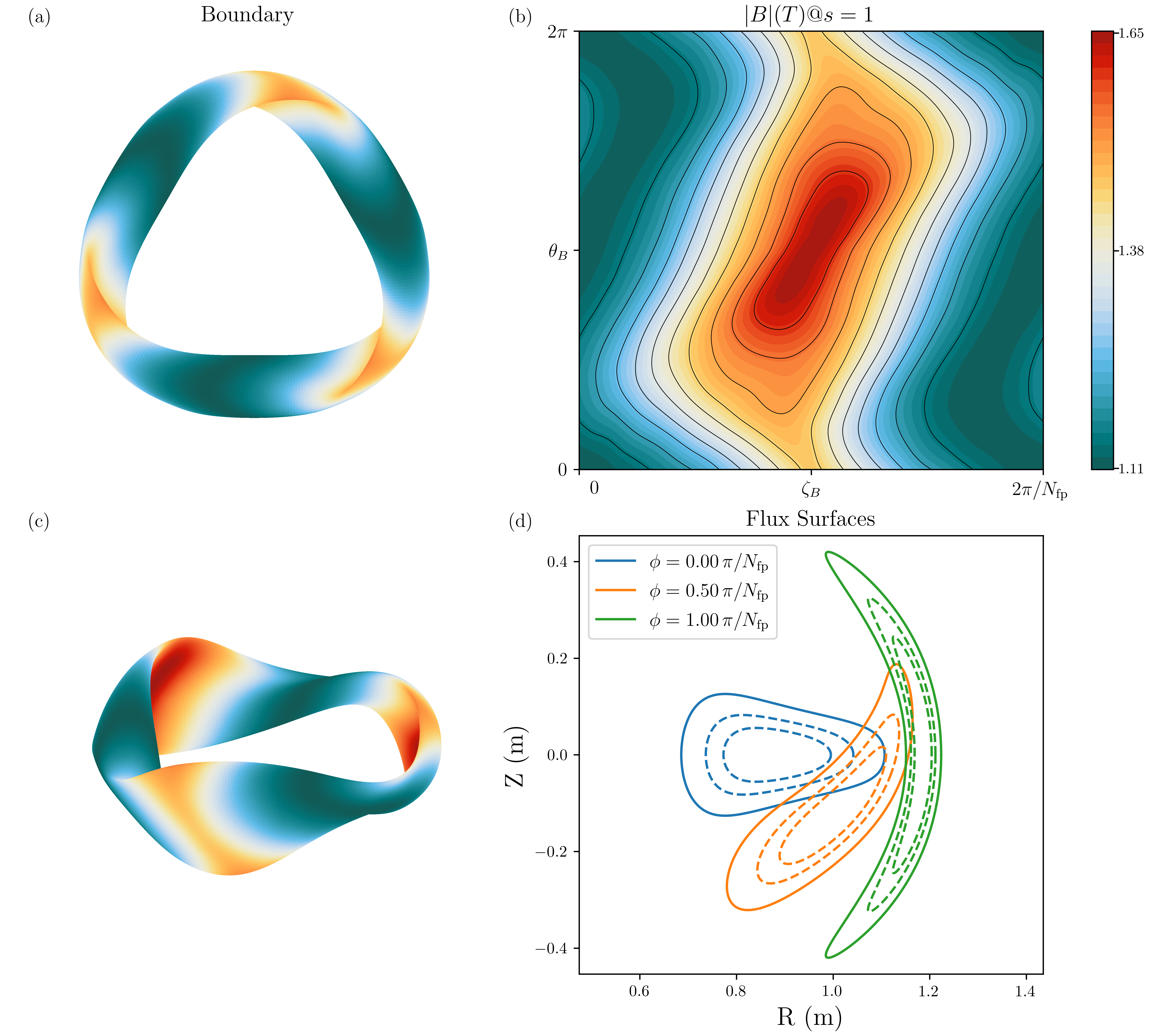}
    \caption{PO-pwO configuration with magnetic well.
            (a) Three-dimensional plot of the equilibrium boundary. 
            (b) Magnetic field strength on the optimized outermost closed flux surface in Boozer coordinates. 
            (c) Magnetic well presented in the equilibrium.
            (d) Cross-sectional views of the equilibrium (dashed lines for internal flux surfaces).
            }
    \label{fig:POpwOWell}
\end{figure}

The optimized configuration is shown in Fig~\ref{fig:POpwOWell}. 
A vacuum magnetic well with depth $\approx 0.023$ appears, as shown in Fig.~\ref{fig:POpwOWell}(c).. 
At the same time, the configuration retains only a small departure from exact omnigenity~\eqref{eq:pJpa} and a low effective ripple $\epsilon_{\text{eff}}^{3/2}$, as shown in Fig.~\ref{fig:pJpa} and Fig.~\ref{fig:eps}, confirming good approximation to PO-pwO. 
At reactor scale, the alpha-particle confinement also remains excellent: the maximum alpha-particle loss over a typical slowing-down time of $0.2~\text{s}$ stays below $2\%$, as shown in Fig.~\ref{fig:loss}. 
This example demonstrates that within the ``squeeze'' framework, one can realize a controlled trade-off between ideal-MHD stability and neoclassical transport by tuning only the mapping parameters, without additional cost functions. 
In this sense, the ``squeeze'' method offers a robust route to families of PO–pwO configurations that simultaneously address transport and stability-related design objectives.

\subsection{$D^*_{11}$ and $D^*_{31}$ coefficients} % or "physics properties"
% Focus on Bootstrap Current
We can further investigate the monoenergetic neoclassical transport coefficients $D^*_{11}$, associated with radial transport, and $D^*_{31}$, related to bootstrap current\cite{beidler_benchmarking_2011}. 
The bootstrap current vanishes in perfectly QI/PO configurations. 
Recent work has proposed the existence of pwO fields in which the bootstrap current is exactly zero\cite{calvo_piecewise_2025}. 
Consequently, when combining PO and pwO concepts, it is essential to preserve this desirable property as much as possible. This has been recently formalized in \cite{Velasco2026}. Even though, within this work, the shape of the pwO region has not yet been controlled for reduced bootstrap current, we will show that the bootstrap current is indeed small in some cases. 
The parallel and radial transport coefficients are computed by \texttt{MONKES} code\cite{escoto_monkes_2024}.
The precise normalization conventions adopted for $D^*_{11}$ and $D^*_{31}$, as well as their relation to the quantities are summarized in ~\ref{app:norm}

\begin{figure*}[t]
    \centering
    \includegraphics[width=1\linewidth]{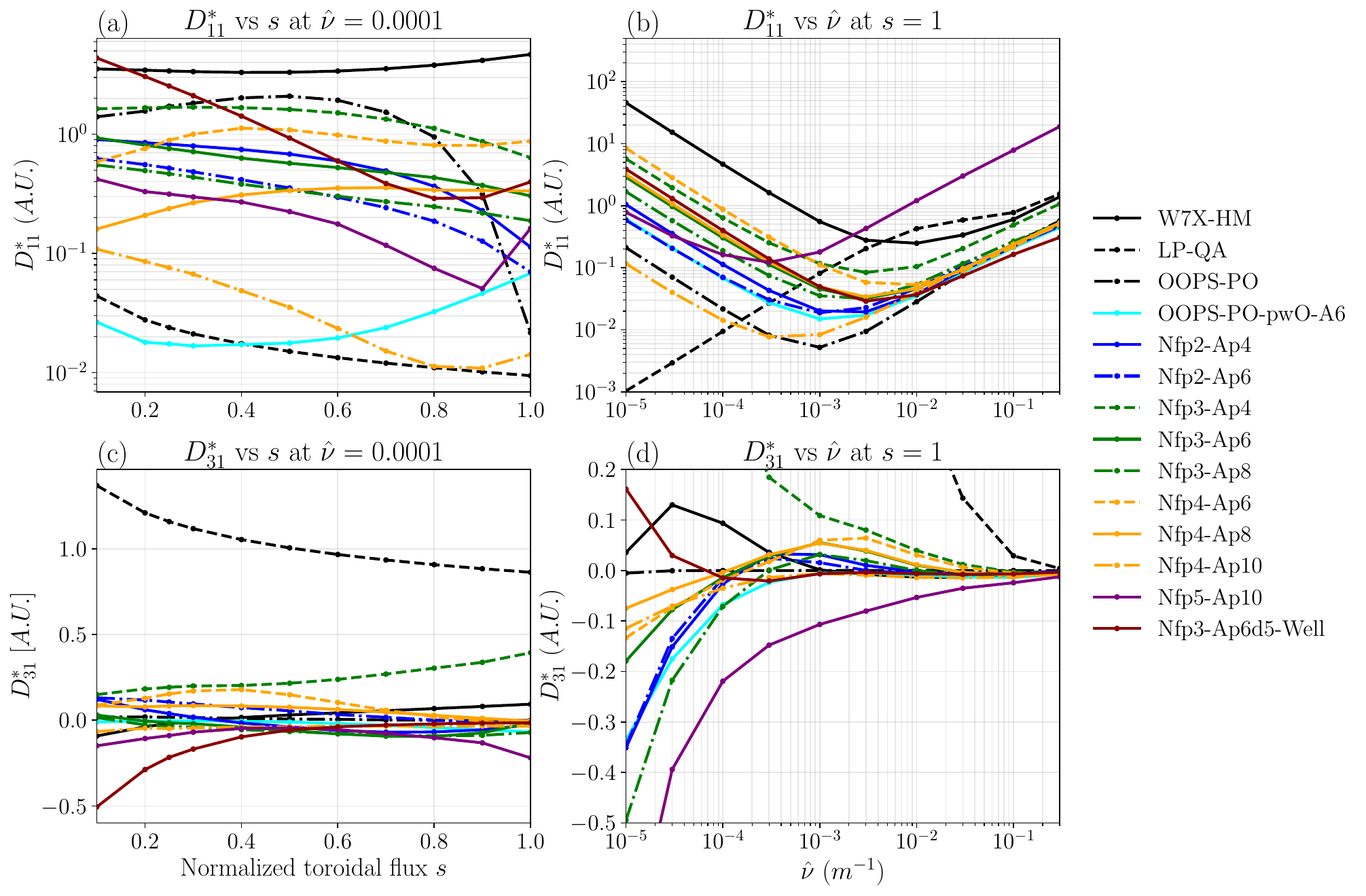}
    \caption{
    Comparisons of normalized transport coefficients for various configurations.
    (a) and (c)$D_{11}^*,D_{31}^*$ vs $s$ at $\hat{\nu}=10^{-4} \ \mathrm{m}^{-1}$.
    (b) and (d)$D_{11}^*,D_{31}^*$ vs $\hat{\nu}$ at $s=1$.
    }
    \label{fig:monkes-D-both}
    
\end{figure*}

%we are able to quickly evaluate the transport coefficients for more than a dozen configurations.
Fig.~\ref{fig:monkes-D-both} presents the values of $D^*_{11}$ and $D^*_{31}$ on the LCFS (the optimized surface) for all  PO-pwO configurations mentioned above, as well as their radial profiles across flux surfaces at a fixed collision frequency $\hat{\nu} = 10^{-5} \ \mathrm{m}^{-1}$.
The W7-X high-mirror (W7X-HM), the precise QA (LP-QA) \cite{landreman_magnetic_2022}, and the PO configuration from \cite{liu_optimizing_2025} (OOPS-PO) are included as a reference.

All PO-pwO configurations exhibit low $D^*_{11}$ (smaller than W7X-HM) coefficients in the $1/\nu$ regime, consistent with the results in the effective ripple.
For configurations with the same $N_\text{fp}$, a larger aspect ratio leads to lower radial transport coefficients.
The distribution of the bootstrap coefficient, $D^*_{31}$, exhibits a wide range.
OOPS-PO has the lowest bootstrap current, since it is a pure omnigenity with high accuracy.
As the only QA configuration, LP-QA has the largest value.
W7X-HM has considerably small bootstrap currents because it was explicitly optimized for low bootstrap currents \cite{Beidler1990}.
Some PO-pwO configurations have $D^*_{31}$ values comparable to W7X-HM in the collisionality regime for fusion plasmas.
Others have relatively large bootstrap coefficients.
This is not surprising since they have departures from an ideal PO-pwO field and we have not enforced the zero-bootstrap-current condition ($w_2=\pi$) in the pwO region.
Within the same $N_\text{fp}$, more compact configurations tend to have larger bootstrap currents.

\subsection{W7-X as a near PO-pwO configuration}
\begin{figure}
    \centering
    \includegraphics[width=1\linewidth]{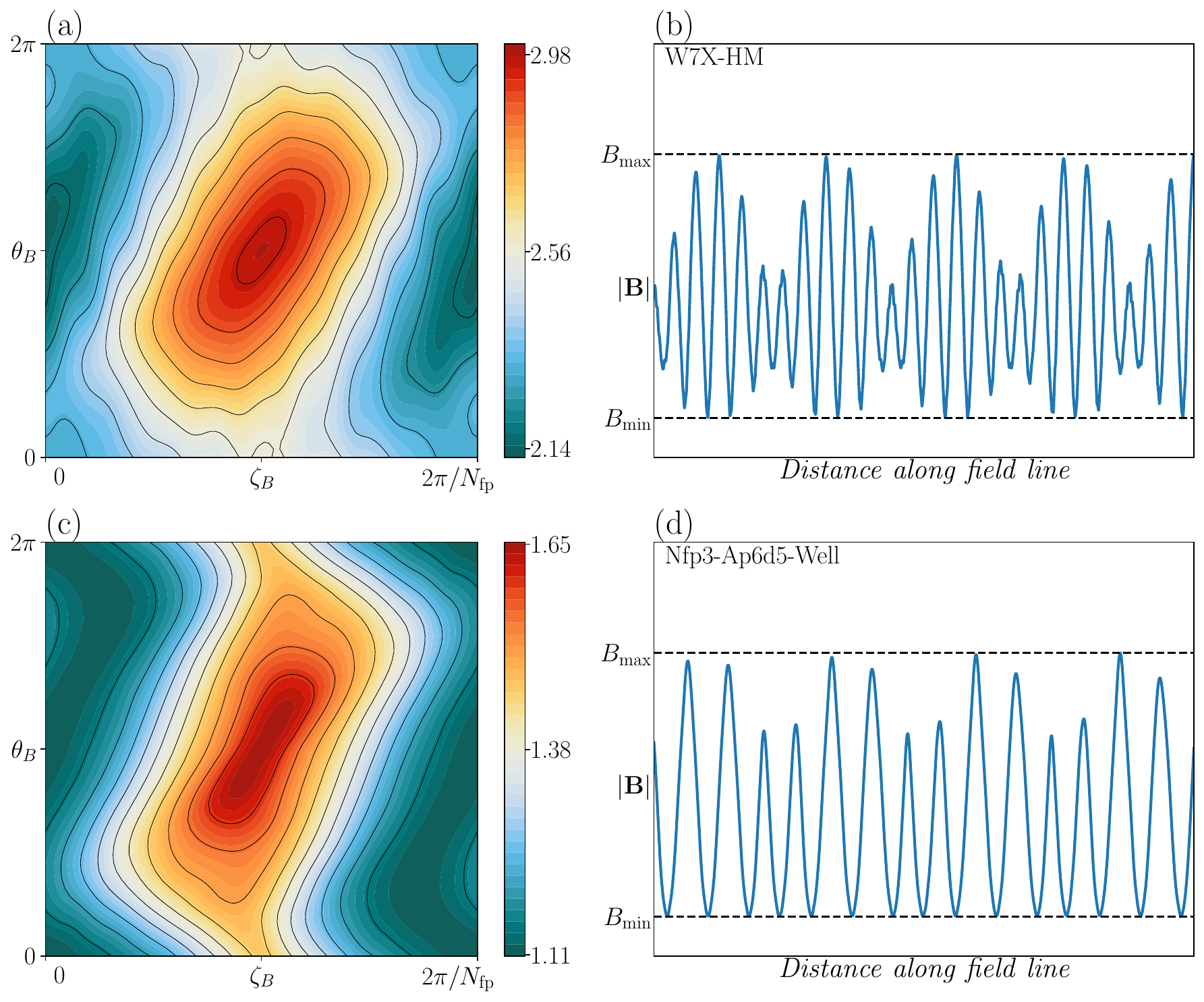}
    \caption{Magnetic-field-strength distribution in Boozer coordinates and along field lines of the W7X-HM and Nfp3-Ap6d5-Well configurations. Panels (a,b) correspond to W7X-HM on an inner flux surface ($s\approx0.5$), and panels (c,d) to  Nfp3-Ap6d5-Well on LCFS.
    \label{fig:w7x}}
\end{figure}

It was argued that a smooth pwO field resembles W7-X standard and high-mirror \cite{velasco_exploration_2025, Velasco2026}.
The $B$ contours of PO-pwO fields (as in Fig.~\ref{fig:all-boozer}) share a lot of similarities with W7-X.
To make this connection more explicit, we further compare the optimized $N_\text{fp} = 3$, $A_p = 6.5$ configuration (Nfp3-Ap6d5-Well) with W7-X (high-mirror), as shown in Fig.~\ref{fig:w7x}.
Although W7X-HM is a fixed-boundary equilibrium, the boundary comes from free-boundary reconstructions using as-built coils. 
Therefore, coil ripples are intrinsically included and this is confirmed in the $B$ contours.
To reduce the influence of this discrete-coil ripple in the comparison, for W7X-HM we use an inner flux surface with \texttt{ns}=50, corresponding approximately to $s=0.5$, rather than the LCFS. 
%A physically realistic magnetic-field distribution on the LCFS generally contains corrugations produced by the discrete modular coils, so perfectly smooth $B$-contours are not expected. 
Even with this corrugation, these two cases share similar structures.
In particular, they both show a low-field region that remains broadly consistent with PO-like contours (less so in the case of W7-X, where a local minimum of $B$ exists), together with a high-field region containing localized closed $B_{\max}$ domains.
Field strength distributions along a field line are plotted in Fig.~\ref{fig:w7x}.
Multiple local wells along the field line are observed, as a result of locally-closed $B$ contours.
This differs from exact omnigenity, which typically has a single maximum/minimum.
However, the minima of B are equally spaced and, specially in the case of Nfp3-Ap6d5-Well, have the same value, similar to the concept reported by Mynick et al. \cite{mynick}, enabling good confinement for trapped particles.
W7-X is an approximation to PO-pwO and the deviation is probably a compromise of other optimization targets, like MHD stability, coil simplification, etc.

\section{Summary and discussion}\label{sec:summary}
In this work, we introduce a method for optimizing stellarator configurations that combine PO and pwO.
The construction is carried out with the \texttt{OOPS} method through a “squeeze” process that compresses the $B_{\max}$ region to create an approximately pwO domain while preserving omnigenity in other regions.
The PO-pwO configurations have low neoclassical transport without enforcing strict poloidal closure of $B$ contours in the high-field region.
The extra freedom expands the available design space for future stellarator reactors. 
They are also compatible with the requirement of zero/low bootstrap currents and MHD stability.

% a family of mappings generated through a “squeezing” process that compresses the $B_{\max}$ region to create an approximately closed domain. 
% This construction bridges the original pwO mapping with the Cary–Shasharina-like omnigenity mapping, and can be efficiently targeted within \texttt{OOPS} framework, we enable an efficient optimization of configurations that combine key features of PO and pwO.
% % The resulting configurations retain PO properties over most of low-field while producing a locally closed high-field domain that is compatible with pwO structure.
% As a result, they achieve a substantial reduction in the neoclassical effective ripple without enforcing strict poloidal closure of the high-$|B|$ contours.
% Collectively, these results show that the ``squeezing'' method provides a practical bridge between PO and pwO: it preserves near-PO organization over the low-field region while enabling a locally closed high-field domain that is compatible with pwO-like structure. 

We identify several possible directions for future work. 
First, we can apply the same method to other concepts, combining pwO with quasisymmetry, toroidal omnigenity, and helical omnigenity.
There are new configurations that are worth exploring.
% systematically explore and optimize HO–pwO and TO–pwO variants to test whether the squeezing framework can enforce piecewise-omnigenous constraints robustly across all symmetry directions and over a wider design space. 
Second, we can modify the mapping function to explicitly control the field structure inside the closed high-field region.
It can be used to shape the $B_{\max}$ region closer to an ideal single-valued parallelogram, or to enforce the zero-bootstrap-current condition of $w_2=\pi$.
Third, we can incorporate the method into stellarator reactor designs, where multiple metrics, such as turbulence transport, coil complexity, etc., are targeted simultaneously. 

\section*{Acknowledgments}
This work was supported by the National Natural Science Foundation of China (NSFC) with Grant No. 12475229 and the Strategic Priority Research Program of the Chinese Academy of Sciences with Grant No. XDB0790302. This research was partially supported by grants PID2021-123175NB-I00 and PID2024-155558OB-I00, Ministerio de Ciencia, Innovaci\'on y Universidades, Spain. 

\appendix
\section{Normalization of the monoenergetic coefficients}
\label{app:norm}

The monoenergetic coefficients $\hat{D}_{11}$ and $\hat{D}_{31}$ are related to their normalized forms $D^*_{11}$ and $D^*_{31}$ by
$$
\begin{aligned}
    D^*_{11} &= \frac{8RB_0\iota}{\pi}K_{11}\hat{D}_{11} \\
    D^*_{31} &= \iota B_0 \sqrt{\frac{r_{\text{lcfs}}}{R}} K_{31}\hat{D}_{31} 
\end{aligned}
$$
where $R$ and $r_{\text{lcfs}}$ denote the major and minor radii of the device, $B_0$ is a reference value for $B$ on the flux surface and $\iota$ is the rotational transform. 
The factors $K_{ij}$ serve as normalization factors to change from the radial coordinate $\psi$ to $r$.
In \texttt{MONKES} code output, the quantities labeled $D_{ij}$ correspond to $K_{ij}\hat{D}_{ij}$.

\section*{References}
\bibliographystyle{unsrt}
\bibliography{ref}

@article{cary_helical_1997,
  title = {Helical {{Plasma Confinement Devices}} with {{Good Confinement Properties}}},
  author = {Cary, John R. and Shasharina, Svetlana G.},
  year = {1997},
  month = jan,
  journal = {Physical Review Letters},
  volume = {78},
  number = {4},
  pages = {674--677},
  publisher = {American Physical Society},
  doi = {10.1103/PhysRevLett.78.674},
  urldate = {2024-10-07}
}

@article{calvo_piecewise_2025,
  title = {Piecewise Omnigenous Stellarators with Zero Bootstrap Current},
  author = {Calvo, Iv{\'a}n and Velasco, Jos{\'e} Luis and Helander, Per and Parra, F{\'e}lix I.},
  year = {2025},
  month = aug,
  journal = {Physical Review E},
  volume = {112},
  number = {2},
  pages = {L023201},
  publisher = {American Physical Society},
  doi = {10.1103/tnh1-mq88},
  urldate = {2025-08-17}
}

@article{escoto_evaluation_2025,
  title = {Evaluation of Neoclassical Transport in Nearly Quasi-Isodynamic Stellarator Magnetic Fields Using {{MONKES}}},
  author = {Escoto, F.J. and Velasco, J.L. and Calvo, I. and S{\'a}nchez, E.},
  year = {2025},
  month = feb,
  journal = {Nuclear Fusion},
  volume = {65},
  number = {3},
  pages = {036017},
  publisher = {IOP Publishing},
  issn = {0029-5515},
  doi = {10.1088/1741-4326/ada7e0},
  urldate = {2025-08-17},
  langid = {english}
}

@article{velasco_exploration_2025,
  title = {Exploration of the Parameter Space of Piecewise Omnigenous Stellarator Magnetic Fields},
  author = {Velasco, J.L. and S{\'a}nchez, E. and Calvo, I.},
  year = {2025},
  month = apr,
  journal = {Nuclear Fusion},
  volume = {65},
  number = {5},
  pages = {056012},
  publisher = {IOP Publishing},
  issn = {0029-5515},
  doi = {10.1088/1741-4326/adc4f6},
  urldate = {2025-08-17},
  langid = {english}
}

@article{velasco_piecewise_2024,
  title = {Piecewise {{Omnigenous Stellarators}}},
  author = {Velasco, J. L. and Calvo, I. and Escoto, F. J. and S{\'a}nchez, E. and Thienpondt, H. and Parra, F. I.},
  year = {2024},
  month = oct,
  journal = {Physical Review Letters},
  volume = {133},
  number = {18},
  pages = {185101},
  publisher = {American Physical Society},
  doi = {10.1103/PhysRevLett.133.185101},
  urldate = {2024-11-09},
  langid = {american}
}

@article{landreman_magnetic_2022,
  title = {Magnetic {{Fields}} with {{Precise Quasisymmetry}} for {{Plasma Confinement}}},
  author = {Landreman, Matt and Paul, Elizabeth},
  year = {2022},
  month = jan,
  journal = {Physical Review Letters},
  volume = {128},
  number = {3},
  pages = {035001},
  publisher = {American Physical Society},
  doi = {10.1103/PhysRevLett.128.035001},
  urldate = {2024-10-14}
}

@article{dudtMagneticFieldsGeneral2024,
  title = {Magnetic Fields with General Omnigenity},
  author = {Dudt, Daniel W. and Goodman, Alan G. and Conlin, Rory and Panici, Dario and Kolemen, Egemen},
  year = {2024},
  month = feb,
  journal = {Journal of Plasma Physics},
  volume = {90},
  number = {1},
  pages = {905900120},
  issn = {0022-3778, 1469-7807},
  doi = {10.1017/S0022377824000151},
  urldate = {2024-09-13},
  langid = {english}
}

@article{goodman_constructing_2023,
  title = {Constructing Precisely Quasi-Isodynamic Magnetic Fields},
  author = {Goodman, A. G. and Mata, K. Camacho and Henneberg, S. A. and Jorge, R. and Landreman, M. and Plunk, G. G. and Smith, H. M. and Mackenbach, R. J. J. and Beidler, C. D. and Helander, P.},
  year = {2023},
  month = oct,
  journal = {Journal of Plasma Physics},
  volume = {89},
  number = {5},
  pages = {905890504},
  issn = {0022-3778, 1469-7807},
  doi = {10.1017/S002237782300065X},
  urldate = {2024-10-07},
  langid = {english}
}

@article{goodman_quasi-isodynamic_2024,
  title = {Quasi-{{Isodynamic Stellarators}} with {{Low Turbulence}} as {{Fusion Reactor Candidates}}},
  author = {Goodman, Alan G. and Xanthopoulos, Pavlos and Plunk, Gabriel G. and Smith, H{\aa}kan and N{\"u}hrenberg, Carolin and Beidler, Craig D. and Henneberg, Sophia A. and {Roberg-Clark}, Gareth and Drevlak, Michael and Helander, Per},
  year = {2024},
  month = jun,
  journal = {PRX Energy},
  volume = {3},
  number = {2},
  pages = {023010},
  publisher = {American Physical Society},
  doi = {10.1103/PRXEnergy.3.023010},
  urldate = {2024-10-07},
  langid = {american}
}

@article{liu_optimizing_2025,
  author  = {Hengqian Liu and Guodong Yu and Caoxiang Zhu and Ge Zhuang},
  title   = {Optimizing omnigenity like quasisymmetry for stellarators},
  year    = {2025},
  journal = {arXiv:2502.09350},
}

@article{helander_bootstrap_2009,
  title = {Bootstrap Current and Neoclassical Transport in Quasi-Isodynamic Stellarators},
  author = {Helander, P. and N{\"u}hrenberg, J.},
  year = {2009},
  month = feb,
  journal = {Plasma Physics and Controlled Fusion},
  volume = {51},
  number = {5},
  pages = {055004},
  issn = {0741-3335},
  doi = {10.1088/0741-3335/51/5/055004},
  urldate = {2024-10-22},
  langid = {english}
}

@misc{cadena_constellaration_2025,
  title = {{{ConStellaration}}: {{A}} Dataset of {{QI-like}} Stellarator Plasma Boundaries and Optimization Benchmarks},
  shorttitle = {{{ConStellaration}}},
  author = {Cadena, Santiago A. and Merlo, Andrea and Laude, Emanuel and Bauer, Alexander and Agrawal, Atul and Pascu, Maria and Savtchouk, Marija and Guiraud, Enrico and Bonauer, Lukas and Hudson, Stuart and Kaiser, Markus},
  year = {2025},
  month = jun,
  number = {arXiv:2506.19583},
  eprint = {2506.19583},
  primaryclass = {cs},
  publisher = {arXiv},
  doi = {10.48550/arXiv.2506.19583},
  urldate = {2025-07-22},
  archiveprefix = {arXiv},
  langid = {american}
}

@article{rodriguez_near-axis_2025,
  title = {Near-Axis Description of Stellarator-Symmetric Quasi-Isodynamic Stellarators to Second Order},
  author = {Rodr{\'i}guez, E. and Plunk, G. G. and Jorge, R.},
  year = {2025},
  month = apr,
  journal = {Journal of Plasma Physics},
  volume = {91},
  number = {2},
  pages = {E59},
  issn = {0022-3778, 1469-7807},
  doi = {10.1017/S0022377825000157},
  urldate = {2025-08-17},
  langid = {english}
}

@misc{gaur_omnigenous_2025,
  title = {Omnigenous Umbilic Stellarators},
  author = {Gaur, R. and Panici, D. and Elder, T. M. and Landreman, M. and Unalmis, K. E. and Elmacioglu, Y. and Dudt, D. and Conlin, R. and Kolemen, E.},
  year = {2025},
  month = may,
  number = {arXiv:2505.04211},
  eprint = {2505.04211},
  primaryclass = {physics},
  publisher = {arXiv},
  doi = {10.48550/arXiv.2505.04211},
  urldate = {2025-08-17},
  archiveprefix = {arXiv}
}

@mastersthesis{fernandez_pacheco_albandea_piecewise_nodate,
  title = {Piecewise Omnigenous Stellarator Configurations Optimized  with Respect to Reactor-Relevant Physics Criteria},
  author = {Fern{\'a}ndez Pacheco Albandea, V{\'i}ctor}
}

@article{landremanOmnigenityGeneralizedQuasisymmetrya2012,
  title = {Omnigenity as Generalized Quasisymmetry)},
  author = {Landreman, Matt and Catto, Peter J.},
  year = {2012},
  month = mar,
  journal = {Physics of Plasmas},
  volume = {19},
  number = {5},
  pages = {056103},
  issn = {1070-664X},
  doi = {10.1063/1.3693187},
  urldate = {2024-09-14},
  langid = {american}
}

@article{parra_less_2015,
  title = {Less Constrained Omnigeneous Stellarators},
  author = {Parra, Felix I. and Calvo, Iv{\'a}n and Helander, Per and Landreman, Matt},
  year = {2015},
  month = feb,
  journal = {Nuclear Fusion},
  volume = {55},
  number = {3},
  pages = {033005},
  publisher = {IOP Publishing},
  issn = {0029-5515},
  doi = {10.1088/0029-5515/55/3/033005},
  urldate = {2024-10-21},
  langid = {english},
}

@inproceedings{okamuramagnetic,
  author    = {Okamura, Shoichi},
  title     = {Magnetic coil design for the improved configuration of LHD},
  booktitle = {Proceedings of the 41st EPS Conference on Plasma Physics},
  address   = {Berlin, Germany},
  year      = {2014},
  month     = jun,
  note      = {P4.011},
  url       = {https://info.fusion.ciemat.es/OCS/EPS2014PAP/pdf/P4.011.pdf},
}

@article{kruskal_equilibrium_1958,
    title = {Equilibrium of a {Magnetically} {Confined} {Plasma} in a {Toroid}},
    volume = {1},
    issn = {0031-9171},
    url = {https://doi.org/10.1063/1.1705884},
    doi = {10.1063/1.1705884},
    number = {4},
    urldate = {2025-12-09},
    journal = {The Physics of Fluids},
    author = {Kruskal, M. D. and Kulsrud, R. M.},
    month = jul,
    year = {1958},
    pages = {265--274},
}

@article{boozer_required_2023,
    title = {Required toroidal confinement for fusion and omnigeneity},
    volume = {30},
    issn = {1070-664X},
    url = {https://doi.org/10.1063/5.0147120},
    doi = {10.1063/5.0147120},
    language = {en-US},
    number = {6},
    urldate = {2024-09-13},
    journal = {Physics of Plasmas},
    author = {Boozer, Allen H.},
    month = jun,
    year = {2023},
    pages = {062503},
}

@article{najmabadi_aries-cs_2008,
    title = {The {ARIES}-{CS} {Compact} {Stellarator} {Fusion} {Power} {Plant}},
    volume = {54},
    issn = {1536-1055},
    url = {https://doi.org/10.13182/FST54-655},
    doi = {10.13182/FST54-655},
    number = {3},
    urldate = {2025-12-09},
    journal = {Fusion Science and Technology},
    author = {Najmabadi, F. and Raffray, A. R. and Abdel-Khalik, S. I. and Bromberg, L. and Crosatti, L. and El-Guebaly, L. and Garabedian, P. R. and Grossman, A. A. and Henderson, D. and Ibrahim, A. and Ihli, T. and Kaiser, T. B. and Kiedrowski, B. and Ku, L. P. and Lyon, J. F. and Maingi, R. and Malang, S. and Martin, C. and Mau, T. K. and Merrill, B. and Moore, R. L. and Peipert Jr., R. J. and Petti, D. A. and Sadowski, D. L. and Sawan, M. and Schultz, J. H. and Slaybaugh, R. and Slattery, K. T. and Sviatoslavsky, G. and Turnbull, A. and Waganer, L. M. and Wang, X. R. and Weathers, J. B. and Wilson, P. and Waldrop III, J. C. and Yoda, M. and Zarnstorffh, M.},
    month = oct,
    year = {2008},
    note = {Publisher: American Nuclear Society
\_eprint: https://doi.org/10.13182/FST54-655},
    pages = {655--672},
}

@article{albert_symplectic_2020,
    title = {Symplectic integration with non-canonical quadrature for guiding-center orbits in magnetic confinement devices},
    volume = {403},
    issn = {0021-9991},
    url = {https://www.sciencedirect.com/science/article/pii/S0021999119307703},
    doi = {10.1016/j.jcp.2019.109065},
    urldate = {2024-11-09},
    journal = {Journal of Computational Physics},
    author = {Albert, Christopher G. and Kasilov, Sergei V. and Kernbichler, Winfried},
    month = feb,
    year = {2020},
    pages = {109065},
}

@article{albert_accelerated_2020,
    title = {Accelerated methods for direct computation of fusion alpha particle losses within, stellarator optimization},
    volume = {86},
    issn = {0022-3778, 1469-7807},
    url = {https://www.cambridge.org/core/journals/journal-of-plasma-physics/article/accelerated-methods-for-direct-computation-of-fusion-alpha-particle-losses-within-stellarator-optimization/E4775DC624ABE06966649227A0953FBD},
    doi = {10.1017/S0022377820000203},
    language = {en},
    number = {2},
    urldate = {2024-11-09},
    journal = {Journal of Plasma Physics},
    author = {Albert, Christopher G. and Kasilov, Sergei V. and Kernbichler, Winfried},
    month = apr,
    year = {2020},
    pages = {815860201},
}

@article{landreman_simsopt_2021,
    title = {{SIMSOPT}: {A} flexible framework for stellarator optimization},
    volume = {6},
    issn = {2475-9066},
    shorttitle = {{SIMSOPT}},
    url = {https://joss.theoj.org/papers/10.21105/joss.03525},
    doi = {10.21105/joss.03525},
    language = {en},
    number = {65},
    urldate = {2024-10-23},
    journal = {Journal of Open Source Software},
    author = {Landreman, Matt and Medasani, Bharat and Wechsung, Florian and Giuliani, Andrew and Jorge, Rogerio and Zhu, Caoxiang},
    month = sep,
    year = {2021},
    pages = {3525},
}

@article{hirshman_steepestdescent_1983,
    title = {Steepest‐descent moment method for three‐dimensional magnetohydrodynamic equilibria},
    volume = {26},
    issn = {0031-9171},
    url = {https://doi.org/10.1063/1.864116},
    doi = {10.1063/1.864116},
    number = {12},
    urldate = {2024-10-23},
    journal = {The Physics of Fluids},
    author = {Hirshman, S. P. and Whitson, J. C.},
    month = dec,
    year = {1983},
    pages = {3553--3568},
}

@article{greene_brief_1998,
    title = {A {BRIEF} {REVIEW} {OF} {MAGNETIC} {WELLS}},
    language = {en},
    author = {Greene, J M},
    year = {1998},
}

@article{beidler_benchmarking_2011,
    title = {Benchmarking of the mono-energetic transport coefficients—results from the {International} {Collaboration} on {Neoclassical} {Transport} in {Stellarators} ({ICNTS})},
    volume = {51},
    issn = {0029-5515},
    url = {https://doi.org/10.1088/0029-5515/51/7/076001},
    doi = {10.1088/0029-5515/51/7/076001},
    language = {en},
    number = {7},
    urldate = {2025-10-11},
    journal = {Nuclear Fusion},
    author = {Beidler, C.D. and Allmaier, K. and Isaev, M.Yu. and Kasilov, S.V. and Kernbichler, W. and Leitold, G.O. and Maaßberg, H. and Mikkelsen, D.R. and Murakami, S. and Schmidt, M. and Spong, D.A. and Tribaldos, V. and Wakasa, A.},
    month = jun,
    year = {2011},
    pages = {076001},
}

@article{escoto_monkes_2024,
    title = {{MONKES}: a fast neoclassical code for the evaluation of monoenergetic transport coefficients in stellarator plasmas},
    volume = {64},
    issn = {0029-5515},
    shorttitle = {{MONKES}},
    url = {https://doi.org/10.1088/1741-4326/ad3fc9},
    doi = {10.1088/1741-4326/ad3fc9},
    language = {en},
    number = {7},
    urldate = {2025-10-11},
    journal = {Nuclear Fusion},
    author = {Escoto, F.J. and Velasco, J.L. and Calvo, I. and Landreman, M. and Parra, F.I.},
    month = jun,
    year = {2024},
    note = {Publisher: IOP Publishing},
    pages = {076030},
}

@article{nemov_evaluation_1999,
    title = {Evaluation of $1/\nu$ neoclassical transport in stellarators},
    volume = {6},
    issn = {1070-664X},
    url = {https://pubs.aip.org/aip/pop/article/6/12/4622/264236/Evaluation-of-1-neoclassical-transport-in},
    doi = {10.1063/1.873749},
    language = {en},
    number = {12},
    urldate = {2025-06-19},
    journal = {Physics of Plasmas},
    author = {Nemov, V. V. and Kasilov, S. V. and Kernbichler, W. and Heyn, M. F.},
    month = dec,
    year = {1999},
    note = {Publisher: AIP Publishing},
    pages = {4622--4632},
}

@article{sanchez_2023,
doi = {10.1088/1741-4326/accd82},
url = {https://doi.org/10.1088/1741-4326/accd82},
year = {2023},
month = {may},
publisher = {IOP Publishing},
volume = {63},
number = {6},
pages = {066037},
author = {Sánchez, E. and Velasco, J.L. and Calvo, I. and Mulas, S.},
title = {A quasi-isodynamic configuration with good confinement of fast ions at low plasma $\beta$},
journal = {Nuclear Fusion}
}

@misc{sanchez2025ciematqi4xreactorrelevantquasiisodynamicstellarator,
      title={CIEMAT-QI4X: a reactor-relevant quasi-isodynamic stellarator configuration compatible with an island divertor}, 
      author={E. Sánchez and J. L. Velasco and I. Calvo and J. M. García-Regaña and C. Salcuni and J. A. Alonso},
      year={2025},
      eprint={2512.08825},
      archivePrefix={arXiv},
      primaryClass={physics.plasm-ph},
      url={https://arxiv.org/abs/2512.08825}, 
}

@article{LION2025114868,
title = {Stellaris: A high-field quasi-isodynamic stellarator for a prototypical fusion power plant},
journal = {Fusion Engineering and Design},
volume = {214},
pages = {114868},
year = {2025},
issn = {0920-3796},
doi = {https://doi.org/10.1016/j.fusengdes.2025.114868},
url = {https://www.sciencedirect.com/science/article/pii/S0920379625000705},
author = {J. Lion and J.-C. Anglès and L. Bonauer and A. {Bañón Navarro} and S.A. {Cadena Ceron} and R. Davies and M. Drevlak and N. Foppiani and J. Geiger and A. Goodman and W. Guo and E. Guiraud and F. Hernández and S. Henneberg and R. Herrero and C. Hintze and H. Höchter and J. Jelonnek and F. Jenko and R. Jorge and M. Kaiser and M. Kubie and E. {Lascas Neto} and H. Laqua and M. Leoni and J.F. Lobsien and V. Maurin and A. Merlo and D. Middleton-Gear and M. Pascu and G.G. Plunk and N. Riva and M. Savtchouk and F. Sciortino and J. Schilling and J. Shimwell and A. {Di Siena} and R. Slade and T. Stange and T.N. Todd and L. Wegener and F. Wilms and P. Xanthopoulos and M. Zheng}
}

@article{Hegna_Anderson_Andrew_Ayilaran_Bader_Bohm_Mata_Canik_Carbajal_Cerfon_et, 
title={The Infinity Two fusion pilot plant baseline plasma physics design}, 
volume={91}, DOI={10.1017/S0022377825000364}, number={3}, journal={Journal of Plasma Physics}, author={Hegna, C.C. and Anderson, D.T. and Andrew, E.C. and Ayilaran, A. and Bader, A. and Bohm, T.D. and Mata, K. Camacho and Canik, J.M. and Carbajal, L. and Cerfon, A. and et al.}, year={2025}, pages={E76}}

@article{bindel_2023,
doi = {10.1088/1361-6587/acd141},
url = {https://doi.org/10.1088/1361-6587/acd141},
year = {2023},
month = {may},
publisher = {IOP Publishing},
volume = {65},
number = {6},
pages = {065012},
author = {Bindel, David and Landreman, Matt and Padidar, Misha},
title = {Direct Optimization of Fast-Ion Confinement in Stellarators},
journal = {Plasma Physics and Controlled Fusion}
}

@article{mynick,
    author = {Mynick, H. E.},
    title = {Transport optimization in stellarators},
    journal = {Physics of Plasmas},
    volume = {13},
    number = {5},
    pages = {058102},
    year = {2006},
    month = {05},
    issn = {1070-664X},
    doi = {10.1063/1.2177643},
    url = {https://doi.org/10.1063/1.2177643},
    eprint = {https://pubs.aip.org/aip/pop/article-pdf/doi/10.1063/1.2177643/19944177/058102_1_1.2177643.pdf},
}

@article{Velasco2026,
  title = {Combination of quasi-isodynamic and piecewise omnigenous magnetic fields},
  author = {Velasco, Jos{\'e} Luis and Calvo, Iv{\'a}n and Liu, Hengqian and {Fern{\'a}ndez-Pacheco}, Victor and S{\'a}nchez, Edilberto and Yu, Guodong and Zhu, Caoxiang },
  year = 2026,
  journal = {To be submitted},
}

@article{Beidler1990,
  title = {Physics and {{Engineering Design}} for {{Wendelstein VII-X}}},
  author = {Beidler, Craig and Grieger, G{\"u}nter and Herrnegger, Franz and Harmeyer, Ewald and Kisslinger, Johann and Lotz, Wolf and Maassberg, Henning and Merkel, Peter and N{\"u}hrenberg, J{\"u}rgen and Rau, Fritz and Sapper, J{\"o}rg and Sardei, Francesco and Scardovelli, Ruben and Schl{\"u}ter, Arnulf and Wobig, Horst},
  year = 1990,
  month = jan,
  journal = {Fusion Technology},
  volume = {17},
  number = {1},
  pages = {148--168},
  issn = {0748-1896},
  doi = {10.13182/FST90-A29178},
  urldate = {2023-03-02},
  langid = {english},
}

\end{document}